\documentclass[letterpaper,twocolumn,10pt]{article}
\usepackage{usenix-2020-09}
\usepackage[available]{usenixbadges}

\usepackage{tikz}
\usepackage{amsmath}
\usepackage{url}

\usepackage{microtype}
\microtypecontext{spacing=nonfrench}
\usepackage{amssymb}
\usepackage{bm}
\usepackage{makecell}
\usepackage{booktabs}
\usepackage{algorithm}
\usepackage{algpseudocode}

\usepackage{enumitem}
\usepackage{multirow}
\usepackage{subcaption}
\usepackage{xcolor}
\usepackage{tcolorbox}

\newcommand\tool{\textsc{Dormant}}

\def\eg{\emph{e.g.}}
\def\etc{\emph{etc.}}
\def\etal{\textit{et al.}}

\begin{document}

\date{}

\title{\Large \bf \tool: Defending against Pose-driven Human Image Animation}

\author{{\rm Jiachen Zhou$^{1,2}$, Mingsi Wang$^{1,2}$, Tianlin Li$^{3}$, Guozhu Meng$^{1,2,}$\thanks{Corresponding authors.}~, Kai Chen$^{1,2,}$\textcolor{green!80!black}{\footnotemark[1]}}\\
$^1$Institute of Information Engineering, Chinese Academy of Sciences, China\\
$^2$School of Cyber Security, University of Chinese Academy of Sciences, China\\
$^3$Nanyang Technological University, Singapore\\
\{zhoujiachen, wangmingsi, mengguozhu, chenkai\}@iie.ac.cn, tianlin001@e.ntu.edu.sg\\
}

\maketitle

\begin{abstract}
Pose-driven human image animation has achieved tremendous progress, enabling the generation of vivid and realistic human videos from just one single photo. However, it conversely exacerbates the risk of image misuse, as attackers may use one available image to create videos involving politics, violence, and other illegal content. To counter this threat, we propose \tool, a novel protection approach tailored to defend against pose-driven human image animation techniques. \tool~applies protective perturbation to one human image, preserving the visual similarity to the original but resulting in poor-quality video generation. The protective perturbation is optimized to induce misextraction of appearance features from the image and create incoherence among the generated video frames. Our extensive evaluation across 8 animation methods and 4 datasets demonstrates the superiority of \tool~over 6 baseline protection methods, leading to misaligned identities, visual distortions, noticeable artifacts, and inconsistent frames in the generated videos. Moreover, \tool~shows effectiveness on 6 real-world commercial services, even with fully black-box access.

\textcolor{red}{\textbf{Warning:} This paper contains unfiltered images generated by diffusion models that may be disturbing to some readers.}
\end{abstract}

\section{Introduction}
\label{sec:introduction}

Diffusion models such as Stable Diffusion~\cite{rombach2022high}, DALL$\cdot$E~\cite{ramesh2022hierarchical}, and Imagen~\cite{saharia2022photorealistic} series models, have demonstrated their unprecedented capabilities in the field of image generation. More recently, video diffusion models like Stable Video Diffusion~\cite{blattmann2023stable} and Sora~\cite{openai2024sora}, have also achieved significant advancements, capable of producing movie-quality and professional-grade videos. In particular, pose-driven human image animation methods have emerged~\cite{lei2024comprehensive}, enabling the generation of controllable and realistic human videos by animating reference images according to desired pose sequences. The resulting videos maintain the appearance of the original reference while accurately following the motion guidance provided by the poses. This innovation holds considerable potential across various applications, including social media, entertainment videos, the movie industry, and virtual characters, \etc

While pose-driven human image animation methods have revolutionized video generation, they also significantly lower the barriers to creating deceptive and malicious human videos, raising serious concerns about unauthorized image usage~\cite{pei2024deepfake,liu2024detection,devin2023animate}. With just a single image of the victim, attackers can generate countless human videos, depicting the individual performing any action dictated by the pose input. For example, attackers can create dancing videos of celebrities to post on video platforms for commercial gains, fabricate fake videos of politicians to incite social harm, or produce Not Safe for Work (NSFW) videos that disrupt people's normal lives. These manipulated videos are easy and low-cost to generate, yet they severely violate portrait and privacy rights and potentially cause considerable harm to the unsuspecting individual. 

\begin{figure}[t]
   \centering
   \includegraphics[width=0.464\textwidth]{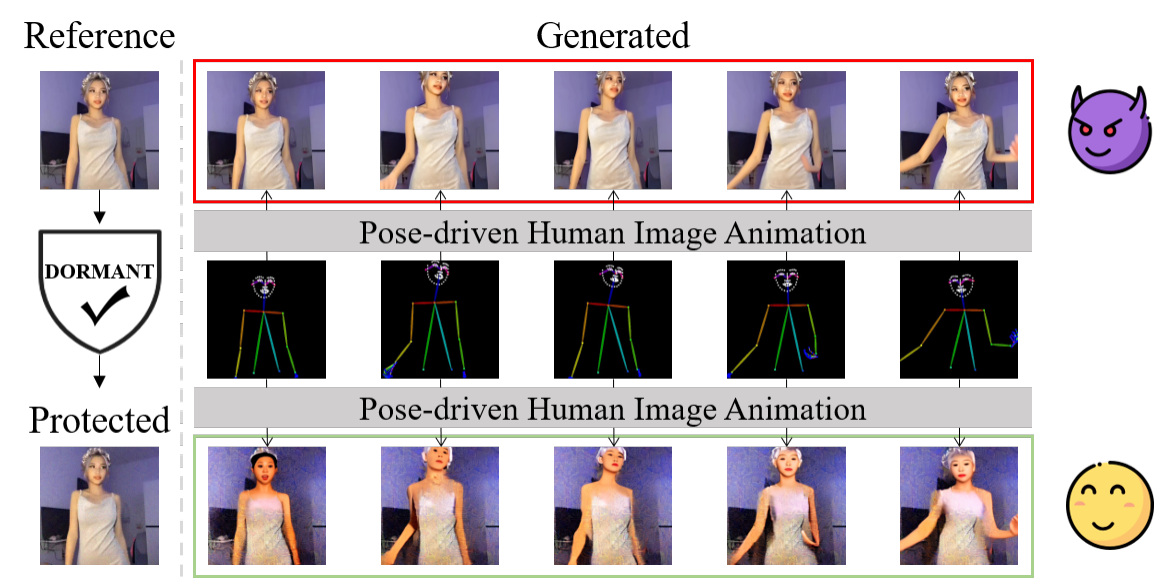}
    \caption{Illustration of defense against pose-driven human image animation. Generated video from the protected image displays mismatched identities and distorted backgrounds.}
    \label{fig:defense against pose-driven human image animation}
\end{figure}

Despite the severity of these hazards, there is a clear lack of research specifically aimed at preventing the misuse of pose-driven human image animation. To our knowledge, the most relevant concurrent work is VGMShield~\cite{pang2024vgmshield}, which is also the only protection method designed to counteract malicious image-to-video generation. VGMShield applies protective perturbation to the image, leading to low-quality video generation by producing incorrect or bizarre frames. The protective perturbation is optimized by targeting the encoding process of Stable Video Diffusion (SVD)~\cite{blattmann2023stable}, deceiving both the image and video encoders into misinterpreting the input image. However, this deviation in the embedding space is not sufficient to effectively disrupt the appearance features contained in images, thus failing to adequately protect human portraits. Furthermore, VGMShield demonstrates limited effectiveness and transferability when defending against non-SVD-based video generation methods, including various pose-driven human image animation techniques. 

While few studies focus specifically on the misuse of image-to-video generation, several protection methods exist to defend against text/image-to-image generation~\cite{liang2023adversarial,liu2024metacloak,salman2023raising,shan2023glaze,van2023anti,xue2024toward,ye2024duaw,zhang2023robustness,zheng2023improving}. These methods typically optimize protective perturbations by targeting the fine-tuning process of customization techniques~\cite{ruiz2023dreambooth}, the encoding process of the Variational AutoEncoder~\cite{esser2021taming}, or the reverse process of the denoising UNet~\cite{ho2020denoising}. However, current protections against image generation are inadequate for defending against pose-driven human image animation, as they fail to effectively deceive the advanced feature extractors in animation methods to disrupt appearance features and break the temporal consistency across frames that is essential in video generation. 

In this paper, we propose \tool~(\underline{D}efending against P\underline{o}se-d\underline{r}iven Hu\underline{m}an Im\underline{a}ge A\underline{n}ima\underline{t}ion), to address this notable gap in effective protection methods that prevent unauthorized image usage in pose-driven human image animation, thereby safeguarding portrait and privacy rights. As illustrated in Figure~\ref{fig:defense against pose-driven human image animation}, \tool~applies protective perturbation to the reference image, resulting in the protected image that appears visually similar to the original but leads to poor-quality video generation when used, causing issues including mismatched appearances, distorted visuals, noticeable artifacts, and incoherent frames. We assume that black-box access to the animation methods and models that the potential attacker might use, and employ a transfer-based strategy to optimize the protective perturbation according to our proposed objective function, utilizing publicly available pre-trained models as surrogates. The optimization objective includes: 1) feature misextraction, which induces deviations in the latent representation and comprehensively disrupts the semantic and fine-grained appearance features of the reference image; and 2) frame incoherence, which targets the appearance alignment and temporal consistency among the generated video frames. We further adopt Learned Perceptual Image Patch Similarity (LPIPS)~\cite{zhang2018unreasonable}, Expectation over Transformation (EoT)~\cite{athalye2018synthesizing}, and Momentum~\cite{dong2018boosting} to enhance the imperceptibility, robustness, and transferability of the protective perturbation, respectively.

We evaluate the protection performance of \tool~on 8 cutting-edge pose-driven human image animation methods. The experimental results across 6 image-level and video-level generative metrics demonstrate the superior effectiveness of \tool, compared to 6 baseline protections against image and video generation. We conduct experiments on 4 datasets, covering tasks such as human dance generation, fashion video synthesis, and speech video generation. Results of human and GPT-4o~\cite{openai2024gpt-4o} studies further highlight the superiority of \tool~from the perspectives of human perception and multimodal large language model. \tool~also exhibits remarkable robustness against 5 popular transformations and 6 advanced purifications. To further assess the transferability of \tool, we conduct additional experiments on 4 image-to-image techniques, 4 image-to-video techniques, and 6 real-world commercial services, \tool~effectively defends against these various generative methods.  

\noindent\textbf{Contributions.} We make the following contributions:

\begin{itemize} [leftmargin=*]
    \item We address the challenge of effectively preventing unauthorized image usage in pose-driven human image animation, safeguarding individuals' rights to portrait and privacy. 
    \item We design novel objective functions to optimize the protective perturbation, inducing misextraction of appearance features and incoherence among generated video frames.
    \item Extensive experiments conducted on 8 pose-driven human image animation methods, 4 image-to-image methods, 4 image-to-video methods, and 6 commercial services highlight the effectiveness and transferability of \tool.
\end{itemize}
\section{Background}
\label{sec:background}

\subsection{Latent Diffusion Model}
\label{subsec:latent diffusion model}

Different from the Denoising Diffusion Probabilistic Model (DDPM)~\cite{ho2020denoising} which directly operates in pixel space, to reduce the computational demand of Diffusion Model (DM), the Latent Diffusion Model (LDM)~\cite{rombach2022high} applies DM training and sampling in the latent space, thus striking a balance between image quality and throughout. LDM utilizes the Variational AutoEncoder (VAE)~\cite{esser2021taming} to provide the low-dimensional latent space, which consists of an encoder $\mathcal{E}$ and a decoder $\mathcal{D}$. More precisely, given an image $x$, the encoder $\mathcal{E}$ maps it to a latent representation: $z=\mathcal{E}(x)$, and then the decoder $\mathcal{D}$ reconstructs it back to the pixel: $\tilde{x}=\mathcal{D}(z)=\mathcal{D}(\mathcal{E}(x))$.

LDM performs the diffusion process proposed in DDPM within the latent space, where the forward process iteratively adds Gaussian noise to the latent representation to make it noisy, while the reverse process predicts and denoises applied noise using a denoising UNet~\cite{ronneberger2015u}. Specifically, given an image $x$, during the forward process, its latent representation $z_{0}=\mathcal{E}(x)$ is perturbed with Gaussian noise over $T$ timesteps, transforming $z_0$ into a standard Gaussian noise $z_T$:
\begin{equation}
\begin{gathered}
    q(z_{1:T}|z_0)=\prod^{T}_{t=1}q(z_t|z_{t-1}), \\ 
    q(z_t|z_{t-1})=\mathcal{N}(z_t;\sqrt{1-\beta_t}z_{t-1},\beta_t\bm{I}),
\end{gathered}
\label{eq:forward process}
\end{equation}
where $\{\beta_t\in(0,1)\}^T_{t=1}$ is the variance schedule. By defining $\alpha_t=1-\beta_t$ and $\overline{\alpha}_t=\prod^t_{i=1}\alpha_i$, we can use the reparameterization technique~\cite{kingma2014auto} to sample $z_t$ at timestep $t$ as follows:
\begin{equation}
\begin{gathered}	   
    q(z_t|z_0)=\mathcal{N}(z_t;\sqrt{\overline{\alpha}_t}z_0,(1-\overline{\alpha}_t)\bm{I}), \\ 
    z_t=\sqrt{\overline{\alpha}_t}z_0+\sqrt{1-\overline{\alpha}_t}\epsilon_t, \quad \epsilon_t\sim\mathcal{N}(0,\bm{I}).
\end{gathered}
\label{eq:reparameterization}
\end{equation}

The reverse process aims to denoise the noisy latent representation $z_T$ back to its original state with the following transition starting at $p(z_T)=\mathcal{N}(z_T;0;\bm{I})$:
\begin{equation}
\begin{gathered}	   
    p_{\theta}(z_{0:T})=p(z_T)\prod^{T}_{t=1}p_{\theta}(z_{t-1}|z_t), \\ p_{\theta}(z_{t-1}|z_t)=\mathcal{N}(z_{t-1};\mu_{\theta}(z_t,c,t),\Sigma_\theta(z_t,c,t)),
\end{gathered}
\label{eq:reverse process}
\end{equation}
where $c$ represents the embedding of conditional information such as text embedding obtained from CLIP ViT-L/14 text encoder~\cite{radford2021learning}, the mean $\mu_{\theta}(z_t,c,t)$ and the variance $\Sigma_\theta(z_t,c,t)$ are computed by the denoising UNet $\epsilon_{\theta}$ parameterized by $\theta$. With Eq. (\ref{eq:reparameterization}), Ho~\etal~\cite{ho2020denoising} simplify the learning objective of the denoising UNet $\epsilon_{\theta}$ in the $\epsilon$-prediction form, where $\epsilon_{\theta}$ is trained to predict the added noise $\epsilon_t$ at timestep $t$:
\begin{equation}
\mathcal{L}_{LDM} = \mathbb{E}_{z_0, c, \epsilon_t \sim \mathcal{N}(0, 1),  t}\Big[ \left\| \epsilon_t - \epsilon_{\theta}(z_{t}, c, t) \right\|_{2}^{2}\Big].
\label{eq:ldm loss}
\end{equation}

Once trained, given $\tilde{z}_T$ sampled from the Gaussian distribution, the denoising UNet $\epsilon_{\theta}$ can be utilized to predict $\epsilon_T$ and compute $\tilde{z}_{T-1}$ according to Eq. (\ref{eq:reverse process}). Subsequently, $\tilde{z}_{T-2}, \tilde{z}_{T-3}, \dots, \tilde{z}_1, \tilde{z}_0$ are progressively calculated, and finally $\tilde{z}_0$ is decoded by $\mathcal{D}$ to generate the image $\tilde{x}=\mathcal{D}(\tilde{z}_0)$.

\subsection{LDM for Human Image Animation}
\label{subsec:LDM for human image animation}

Pose-driven human image animation~\cite{lei2024comprehensive,xu2024magicanimate,hu2024animate,zhu2024champ,wang2024unianimate,wang2024vividpose} is an image-to-video task aimed at generating a human video by animating a static human image based on a user-defined pose sequence. As shown in Figure~\ref{fig:defense against pose-driven human image animation}, the created video needs to accurately align the appearance details from the reference image while following the motion guidance from the pose sequence, which can be produced by DWPose~\cite{yang2023effective} or DensePose~\cite{guler2018densepose}. Recently, LDM-based animation methods have demonstrated superior effectiveness in addressing challenges in appearance alignment and pose guidance, enabling the generation of realistic and coherent videos. 

Using LDM as the backbone, Stable Diffusion (SD) has achieved remarkable success in text/image-to-image generation tasks~\cite{rombach2022high,zhang2023adding}. Pre-trained SD models effectively capture high-quality content priors. However, their network architecture is inherently designed for image generation, lacking the capability to handle the temporal dimension required for video generation. To extend foundational text/image-to-image models for image animation, many works~\cite{xu2024magicanimate,hu2024animate,zhu2024champ,chang2024magicpose,wang2024disco} have incorporated temporal layers~\cite{guo2024animatediff} into the denoising UNet for temporal modeling. Specifically, the feature map $X \in \mathbb{R}^{b \times f \times h \times w \times c}$ is reshaped to $X \in \mathbb{R}^{(b \times h \times w) \times f \times c}$, after which temporal attention is performed to capture the temporal dependencies among frames. Recently, some studies~\cite{peng2024controlnext,zhang2024mimicmotion,wang2024vividpose} have also adopted Stable Video Diffusion~\cite{blattmann2023stable} as the backbone for image animation. SVD is an open-source image-to-video LDM trained on a large-scale video dataset; its strong motion-aware prior knowledge facilitates maintaining temporal consistency in image animation.

While pose-driven human image animation has demonstrated impressive capabilities in creating vivid and realistic human videos, it also raises serious concerns about unauthorized image usage by malicious attackers. With just a single photo of the victim, easily obtained from the web or social media, the malicious attacker can generate an unlimited number of human videos, manipulating the victim to perform any poses and thereby severely violating the individual's rights to publicity and privacy. These manipulated videos could be used for commercial or political purposes, potentially inflicting significant harm on the unsuspecting victim. 

\subsection{Protection Methods against LDM}
\label{subsec:protection methods against LDM}

While few studies addressing the issue of unauthorized image-to-video generation, prior research has proposed various protection methods~\cite{zheng2023improving,van2023anti,liu2024metacloak,shan2023glaze,salman2023raising,ye2024duaw,liang2023adversarial,xue2024toward,zhang2023robustness} against LDM to prevent image misuse in text/image-to-image generation. In the context of image generation, there are two primary LDM-based image misuse scenarios: concept customization and image manipulation. Existing protection methods mainly utilize Projected Gradient Descent (PGD)~\cite{madry2018towards} to generate adversarial examples~\cite{goodfellow2015explaining,carlini2017towards} for LDM, thereby safeguarding against malicious image generation in both misuse scenarios. The defender aims to introduce protective perturbation $\delta$ to the image $x$, such that the protected image $x_p = x + \delta$ can deceive the LDM, resulting in poor-quality generation.

\noindent\textbf{Defending against Concept Customization.} Concept customization techniques~\cite{hu2022lora,gal2023an,ruiz2023dreambooth,kumari2023multi} fine-tune a few parameters of a pre-trained text-to-image model so that it quickly acquires a new concept given only 3-5 images as reference. The training loss of the most popular work DreamBooth~\cite{ruiz2023dreambooth} introduces a prior preservation loss to $\mathcal{L}_{LDM}$ in Eq. (\ref{eq:ldm loss}):
\begin{multline}
\mathcal{L}_{DB}=
\mathbb{E}_{z_0, c, \epsilon_t, \epsilon_{t^{\prime}}^{\prime}, t, t^{\prime}}\Big[\left\| \epsilon_t - \epsilon_{\theta}(z_{t+1}, c, t) \right\|_{2}^{2}
\\
+\lambda \left\| \epsilon_{t^{\prime}}^{\prime} - \epsilon_{\theta}(z^{\prime}_{{t^{\prime}}+1}, c_{pr}, t^{\prime}) \right\|_{2}^{2} \Big],  
\label{eq:db loss}
\end{multline}
where $z^{\prime}_{{t^{\prime}}+1}$ is the noisy latent representation of the class sample $x^{\prime}$ generated from LDM with prior prompt $c_{pr}$, $\lambda$ controls for the weight of the prior term,  which prevents over-fitting and text-shifting problems. Depending on the conceptual properties, there are two variants: style mimicry~\cite{shan2023glaze} and subject recontextualization~\cite{an2024rethinking}. Style mimicry seeks to generate paintings in an artist's style without consent, while subject recontextualization aims to memorize a subject (\eg, a portrait of a victim) and generate it in a new context. Protection methods defending against concept customization~\cite{zheng2023improving,van2023anti,liu2024metacloak} mainly disrupt the learning process in Eq. (\ref{eq:db loss}) by solving the bi-level optimization: $\delta={arg\,max}_{\delta}\mathcal{L}_{LDM}(\mathcal{E}(x+\delta), \theta^{\prime}), s.t. \theta^{\prime}={arg\,min}_{\theta}\mathcal{L}_{DB}(\mathcal{E}(x+\delta), \theta)$, aimed at finding protective perturbation $\delta$ which degrades the personalization generation ability of DreamBooth, thereby preventing unauthorized concept customization.

\noindent\textbf{Defending against Image Manipulation.} Image Manipulation techniques~\cite{brack2024ledits++,meng2022sdedit,brooks2023instructpix2pix,couairon2023diffedit} leverage pre-trained image-to-image models to directly manipulate a single image, altering its appearance or transferring its style. Protection methods aimed at defending against image manipulation focus on disrupting the encoding process ($x\xrightarrow{}z_0$)~\cite{shan2023glaze,salman2023raising,ye2024duaw} and the reverse process ($z_T\xrightarrow{}z_0$)~\cite{liang2023adversarial,xue2024toward,zhang2023robustness}. The protective perturbation $\delta$ is optimized by fooling the VAE $\delta={arg\,max}_{\delta}\left\| \mathcal{E}(x+\delta)-\mathcal{E}(x)\right\|_{2}^{2}$ and the denoising UNet $\delta={arg\,max}_{\delta}\mathcal{L}_{LDM}(\mathcal{E}(x+\delta))$, thus making the LDM to generate images significantly deviating from original image $x$. 

There are several differences between concept customization, image manipulation, and pose-driven human image animation. First, concept customization and image manipulation are image generation tasks, whereas human image animation is categorized as image-to-video generation. Second, concept customization requires multiple images (typically 3-5), while the other two techniques operate on a single image. Third, concept customization involves fine-tuning the LDM to memorize and generate a new concept. In contrast, image manipulation and human image animation leverage frozen pre-trained LDM directly for image or video generation without additional learning. Fourth, concept customization uses images as training data for fine-tuning, image manipulation directly edits the given image itself, while human image animation extracts appearance features from the reference image to guide video generation, rather than using the image directly.

These various ways of using images necessitate the design of specific objective functions for each case when optimizing the protective perturbation. Existing protections against concept customization and image manipulation mainly target the fine-tuning process, or the encoding and reverse processes, respectively. However, these methods are ineffective at disrupting the appearance features contained in human images. As shown in Figure~\ref{fig:qualitative comparisons with baseline protections} in Section~\ref{subsec:protection performance}, only \tool~effectively induces noticeable mismatches in identity appearance between reference images and generated videos. The reason is that protective perturbations optimized with current objectives have a limited protective effect on various well-trained feature extractors used in animation methods. The ablation study results in Figure~\ref{fig:ablation study on the proposed objectives} in Section~\ref{subsec:ablation study} also emphasize the need for specifically designed objective functions to induce feature misextraction, which existing methods lack. Moreover, they also neglect to break the temporal consistency across frames specifically required in video generation. As a result, current protections against image generation are insufficient for defending against pose-driven human image animation.
 
While many protection methods exist to prevent unauthorized image usage in text/image-to-image generation, there is a notable gap in defending against image-to-video generation. To our knowledge, the only concurrent work addressing this issue is VGMShield~\cite{pang2024vgmshield}, which aims to prevent malicious image-to-video generation using SVD~\cite{blattmann2023stable} by generating adversarial examples. More precisely, it attacks the encoding process of SVD, aiming at deceiving the image encoder and video encoder into misinterpreting the image. However, this deviation specifically in the embedding space of SVD is insufficient to effectively disrupt the appearance features in human images and shows limited effectiveness and transferability when defending against non-SVD-based video generation methods, including various pose-driven human image animation techniques. Overall, the ineffectiveness of existing protections against image and video generation highlights the need for novel protection methods specifically designed to prevent malicious pose-driven human image animation.
\section{Methodology}
\label{sec:methodology}

\begin{figure*}[!t]
\centering
\includegraphics[width=0.91\textwidth]{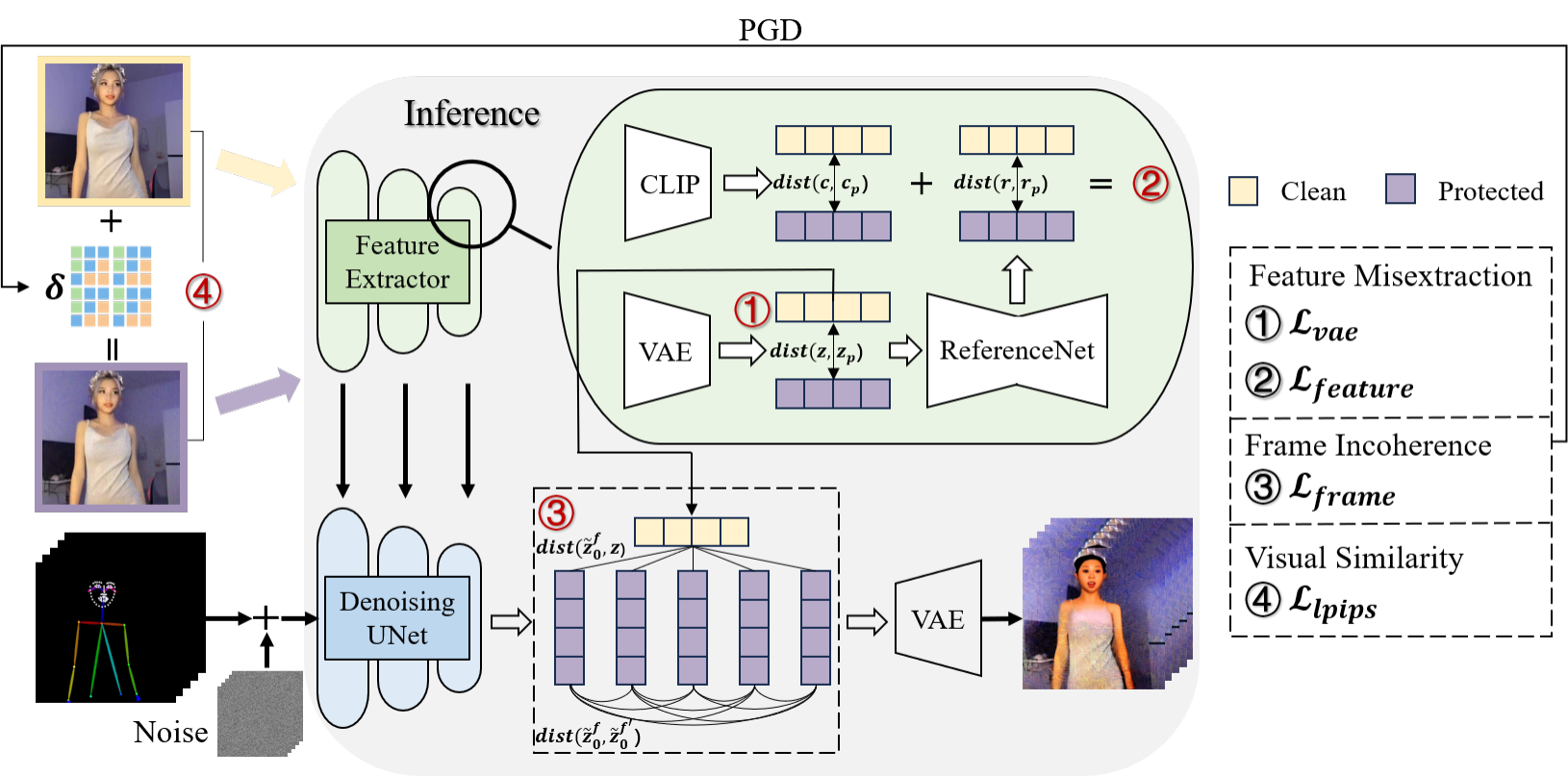}
\caption{Overview of \tool. Here, we present the four components of our proposed objective function $\mathcal{L}_{\tool}$, which includes: $\mathcal{L}_{vae}$ and $\mathcal{L}_{feature}$ for feature misextraction, $\mathcal{L}_{frame}$ for frame incoherence, and $\mathcal{L}_{lpips}$ for visual similarity.}
\label{fig:overview}
\end{figure*}

\tool~prevents image misuse in pose-driven human image animation by adding protective perturbation $\delta$ to the human image $x$, resulting in the protected image $x_p = x + \delta$. $x_p$ is visually similar to $x$ but would lead to poor-quality video generation if used, causing issues such as mismatched appearances, visual distortions, and frame inconsistencies. We assume that a fully black-box setting for the animation methods and models that potential attackers might use and employ a transfer-based strategy to optimize $\delta$ according to our proposed objective function $\mathcal{L}_{\tool}$, relying on some surrogate pre-trained models (\eg, the publicly available CLIP model~\cite{radford2021learning}) to which the defender has white-box access. As shown in Figure~\ref{fig:overview}, $\mathcal{L}_{\tool}$ contains: 1) $\mathcal{L}_{vae}$ and $\mathcal{L}_{feature}$, which aim to induce misextraction of appearance features from the reference image; and 2) $\mathcal{L}_{frame}$, which is designed to cause incoherence among the generated video frames. Additionally, we adopt LPIPS~\cite{zhang2018unreasonable}, EoT~\cite{athalye2018synthesizing}, and Momentum~\cite{dong2018boosting}, to further enhance the imperceptibility, robustness, and transferability of $\delta$, respectively. The protective perturbation is optimized using PGD~\cite{madry2018towards}, which iteratively updates $\delta$ with a step size $\gamma$ under an $L_{\infty}$ norm constraint $\eta$ as follows:
\begin{equation}
    \delta_{i} = \delta_{i-1} + \gamma \cdot \text{sign}(\nabla_\delta \mathcal{L}_{\tool}), \,\,\, s.t. \,\,\, \| \delta \|_\infty \leq \eta.
\label{eq:pgd process}
\end{equation} 

\subsection{Threat Model}
\label{subsec:threat model}

\noindent\textbf{Attacker.} The attacker aims to generate human videos using pose sequences based on images obtained without the owner’s consent. These fabricated videos could be misused for commercial or political purposes, violating the victim's rights to portrait and privacy. We assume that the attacker maliciously acquires multiple images of the victim from the web or social media. The attacker is aware of potential protections and can apply various countermeasures to these images to craft a suitable reference image for video generation. The attacker also has access to various LDMs for pose-driven human image animation, and significant computational power to run them. 

\noindent\textbf{Defender.} We assume that the defender intends to apply protective perturbation to a human image before posting it online or on social media, to prevent potential unauthorized human video generation and thereby safeguard portrait and privacy rights. The protected image appears visually similar to the original image to human perception. However, any videos generated using this protected image would display mismatches in identity appearance and degraded quality, such as distorted backgrounds, visible artifacts, and incoherent frames, \etc~To optimize the protective perturbation, the defender has access to some publicly available feature extractors and LDMs for pose-driven human image animation. However, the defender is unaware of, and therefore unable to approximate, the specific animation methods and models that the attacker might use. The defender could be either the image owner or a trustworthy third party with sufficient computing resources to execute the protection method effectively.

\subsection{Feature Misextraction}
\label{subsec:feature misextraction}

As mentioned in Section~\ref{subsec:protection methods against LDM}, the mainstream usage of the reference image in pose-driven human image animation involves extracting its appearance features to guide video generation. In this context, the objective of optimizing the protective perturbation $\delta$ is to disrupt these appearance features, causing the feature extractor to extract incorrect features from the protected image. As a result, videos generated with this faulty guidance would differ from the reference image, thereby preventing unauthorized usage in human video generation.  

\noindent\textbf{VAE Encoder.} As mentioned in Section~\ref{subsec:latent diffusion model}, to reduce computational demands, LDM first maps the input image to a latent representation using the VAE encoder before performing the diffusion process. Given this, an intuitive strategy for inducing feature misextraction is to disrupt the encoding process, causing the encoder $\mathcal{E}$ to map the protected image $x_p$ to a ``wrong'' latent vector $z_p=\mathcal{E}(x+\delta)$ that deviates significantly from the original latent $z=\mathcal{E}(x)$. Consequently, this misencoded latent $z_p$ would contain incorrect appearance features, thus misleading the video generation process. The optimization objective $\mathcal{L}_{vae}$ is to maximize the distance between $x$ and $x_p$ in the VAE latent space, as defined by: 
\begin{equation}
\underset{\| \delta \|_\infty \leq \eta}{arg\,max} \mathcal{L}_{vae} = \underset{\| \delta \|_\infty \leq \eta}{arg\,max}\left\| \mathcal{E}(x+\delta) - \mathcal{E}(x)\right\|^2_2.
\label{eq:vae loss}
\end{equation}

We utilize the encoder $\mathcal{E}$ from \textit{sd-vae-ft-mse}~\cite{stability2022sd}, a VAE fine-tuned on an enriched dataset with images of humans to improve the reconstruction of faces.  

However, the deviation of the latent representation alone is insufficient to induce a deep-level misextraction of the appearance features. This is because pose-driven human image animation methods typically have stringent requirements for appearance alignment and often use pre-trained or custom-designed models for additional feature extraction. As a result, the attack on the VAE encoder has only a limited effect on these advanced feature extractors. To comprehensively disrupt the appearance features of the image and thus improve the transferability of the protective perturbation against various unknown feature extractors, we further employ CLIP image encoder for semantic feature extraction and ReferenceNet for fine-grained feature extraction during the optimization of $\delta$. 

\noindent\textbf{CLIP Image Encoder.} The CLIP model~\cite{radford2021learning} is trained on a diverse set of text-image pairs, and its image encoder $\mathcal{C}$ can be utilized to extract semantic features from the reference image $x$, which can then be integrated into the Transformer Blocks of the denoising UNet to guide video generation through cross-attention~\cite{wang2024disco}. To ensure that the protected image $x_p = x + \delta$ contains different semantic features from the original image $x$, we further employ the CLIP image encoder $\mathcal{C}$ from \textit{sd-image-variations}~\cite{lambda2022stable} for feature extraction during the optimization of $\delta$. The objective is to maximize the distance between $x$ and $x_p$ in the CLIP embedding space, defined as follows:
\begin{equation}
\underset{\| \delta \|_\infty \leq \eta}{arg\,max}\left\| \mathcal{C}(x+\delta) - \mathcal{C}(x)\right\|^2_2.
\label{eq:clip loss}
\end{equation}

However, CLIP image encoder is known to be less effective at capturing fine-grained details~\cite{xu2024magicanimate,hu2024animate}. There are two main factors for this limitation. First, CLIP is trained to match semantic features between text and images, which are typically sparse and high-level, resulting in a deficit of detailed features. Second, the CLIP image encoder only accepts low-resolution ($224 \times 224$) images as inputs, leading to a loss of fine-grained information. Therefore, relying solely on the CLIP image encoder for optimizing $\delta$ is insufficient, as it mainly extracts semantic features while missing fine-grained details.

\noindent\textbf{ReferenceNet.} Inspired by the recent success of ReferenceNet in detailed feature extraction~\cite{chang2024magicpose,xu2024magicanimate,hu2024animate,zhu2024champ,wang2024vividpose,tong2024musepose,huang2024magicfight}, we further utilize ReferenceNet $\mathcal{R}$ as the feature extractor to optimize $\delta$. ReferenceNet is a copy of the denoising UNet from SD used in image generation\footnote{ReferenceNet is identical to traditional denoising UNet used in image generation, while denoising UNet for human image animation incorporates additional temporal layers to generate video, as mentioned in Section~\ref{subsec:LDM for human image animation}.}, designed to extract appearance features particularly low-level details from the reference human image. Specifically, the feature map $X_1 \in \mathbb{R}^{h \times w \times c}$ from the ReferenceNet is repeated by $f$ times and concatenated with the feature map $X_2 \in \mathbb{R}^{f \times h \times w \times c}$ from the denoising UNet along $w$ dimension. Then spatial attention is performed to transmit the appearance information from the reference image spatially. ReferenceNet inherits weights from the original SD and is further trained on human video frames, enabling it to provide fine-grained and detailed features essential for preserving the reference appearance in video generation. 

In addition to using CLIP image encoder $\mathcal{C}$ to extract semantic features and optimize the protective perturbation $\delta$ according to Eq. (\ref{eq:clip loss}), we also incorporate ReferenceNet $\mathcal{R}$ into the optimization process to enhance fine-grained feature extraction. To ensure that the protected image $x_p$ exhibits distinct detailed features from the perspective of ReferenceNet, we maximize the distance between the corresponding extracted features $\mathcal{R}(\mathcal{E}(x+\delta))$ and $\mathcal{R}(\mathcal{E}(x))$. Moreover, to further improve the transferability of the protective perturbation, we optimize $\delta$ using an ensemble~\cite{dong2018boosting,liu2017delving} of $K=3$ different pre-trained ReferenceNets~\cite{xu2024magicanimate,hu2024animate,chang2024magicpose}. The objective $\mathcal{L}_{feature}$ for feature misextraction using the CLIP image encoder and ReferenceNets is thus defined as follows: 
\begin{equation}
\begin{aligned}
\underset{\| \delta \|_\infty \leq \eta}{arg\,max}\mathcal{L}_{feature} =
\underset{\| \delta \|_\infty \leq \eta}{arg\,max}\left\| \mathcal{C}(x+\delta) - \mathcal{C}(x)\right\|^2_2 \\ + \sum_{k=1}^K\left\| \mathcal{R}_k(\mathcal{E}(x+\delta)) - \mathcal{R}_k(\mathcal{E}(x))\right\|^2_2.
\end{aligned}
\label{eq:feature loss}
\end{equation}

\subsection{Frame Incoherence}
\label{subsec:frame incoherence}

While feature misextraction mainly focuses on disrupting the alignment of reference appearance, it is also crucial to consider the requirement of maintaining temporal consistency across frames in video generation when designing the optimization objective. Based on this, we further propose $\mathcal{L}_{frame}$, which attacks the video generation process of the denoising UNet to induce incoherence among the generated video frames, thus enhancing the protective effect. This requires the defender to provide both the reference image and the pose sequence to simulate the attacker's animation process. However, predicting the exact pose sequences the attacker might use is challenging. To this end, we directly extract the corresponding pose from the reference image and repeat it $F$ times as the pose sequence to guide video generation. 

The protective perturbation $\delta$ can be optimized from two perspectives: 1) maximizing the distance between each video frame and reference image to disrupt appearance alignment; and 2) maximizing the distance between each video frame and others to disrupt video consistency. Since pose-driven human image animation operates in the latent space, these two objectives can be formulated as increasing the distance between the latent vector of the reference image $z = \mathcal{E}(x)$ and that of each generated video frame $\tilde{z}_0^f$, or the distance between the latent vectors of different frames $\tilde{z}_0^f$ and $\tilde{z}_0^{f^\prime}$. The objective $\mathcal{L}_{frame}$ for frame incoherence is defined as follows:
\begin{equation}
\begin{aligned}
\underset{\| \delta \|_\infty \leq \eta}{arg\,max}\mathcal{L}_{frame} = \underset{\| \delta \|_\infty \leq \eta}{arg\,max}\frac{1}{F}\sum_{f=1}^F\left\|\tilde{z}_0^f  - \mathcal{E}(x)\right\|_2^2 \\ 
+ \frac{2}{F(F-1)}\sum_{f=1}^F \sum_{{f^\prime}= f + 1}^F \left\|\tilde{z}_0^f  - \tilde{z}_0^{f^\prime}\right\|_2^2.
\label{eq:frame loss}
\end{aligned}
\end{equation}

We set $F=5$ by default. As mentioned in Section~\ref{subsec:latent diffusion model}, given the total inference steps $T$, we can sample $\tilde{z}_T$ from the Gaussian distribution and compute $\tilde{z}_0$ progressively from timestep $T$ to timestep $0$ according to Eq. (\ref{eq:reverse process}). However, this would result in substantial GPU memory usage and increased time costs. To improve efficiency, we directly estimate $\tilde{z}_0$ using the reparameterization technique outlined in Eq. (\ref{eq:reparameterization}), rather than computing it step by step. Specifically, given a timestep $t$, $\tilde{z}_0$ is estimated by $\tilde{z}_0=(\tilde{z}_t-\sqrt{1-\overline{\alpha}_t}\epsilon_t)/\sqrt{\overline{\alpha}_t}$, where $\overline{\alpha}_t=\prod^t_{i=1}(1-\beta_i$), $\{\beta_t\in(0,1)\}^T_{t=1}$ is the variance schedule, $\epsilon_t$ is predicted by a pre-trained denoising UNet\footnote{\url{https://github.com/MooreThreads/Moore-AnimateAnyone}}, and $\tilde{z}_t$ is sampled from the Gaussian distribution. In each PGD iteration, we randomly sample one timestep $t$ from the last 10 timesteps of $T$ and estimate the corresponding $\tilde{z}_0$.

\subsection{\tool}
\label{subsec:dormant}

Overall, to optimize the protective perturbation $\delta$ for preventing unauthorized image usage in pose-driven human image animation, we propose $\mathcal{L}_{vae}$ and $\mathcal{L}_{feature}$ for feature misextraction, $\mathcal{L}_{frame}$ for frame incoherence. We also apply several techniques to further enhance the perturbation $\delta$ during optimization, including LPIPS~\cite{zhang2018unreasonable} for imperceptibility, EoT~\cite{athalye2018synthesizing} for robustness, and Momentum~\cite{dong2018boosting} for transferability. 

\noindent\textbf{LPIPS.} In addition to constraining the protective perturbation $\delta$ using the $L_\infty$ norm, we also incorporate LPIPS to further improve the imperceptibility of $\delta$. LPIPS utilizes deep features from a pre-trained network to measure image distance, aligning more closely with human perception. Thus we use LPIPS to ensure that the protected image $x_p$ remains visually similar to the original image $x$. Specifically, we include the following regularization term and minimize $\mathcal{L}_{lpips}$ during optimization:
\begin{equation}
\mathcal{L}_{lpips} = \text{max}(\text{LPIPS}(x + \delta, x) - \zeta,0),
\label{eq:lpips loss}
\end{equation}
where $\zeta$ denotes the budget of $\mathcal{L}_{lpips}$, and we set $\zeta=0.1$ by default. During the optimization of $\delta$, if the perceptual distance between $x_p$ and $x$ exceeds the bound $\zeta$, $\mathcal{L}_{lpips}$ introduces a penalty to maintain visual similarity; otherwise, $\mathcal{L}_{lpips}=0$.

\noindent\textbf{EoT.} To enhance the robustness of $\delta$ against various transformations, we adopt EoT into the PGD process. EoT introduces a distribution of transformation functions to the input and optimizes the expectation of the objective function values over these transformations. To reduce time and computational costs, during each PGD step, we randomly sample one of the following transformations and apply it to the protected image $x_p$: Gaussian blur, JPEG compression, Gaussian noise, Random resize (resizing to a random resolution and then back to the original size), or no transformation. These selected transformations are commonly adopted as potential countermeasures in prior studies and have been shown to weaken the effect of protective perturbations~\cite{zhao2024can,an2024rethinking,honig2025adversarial}.

\noindent\textbf{Momentum.} Momentum enhances the transferability of the perturbation by accumulating a velocity vector in the gradient direction of the loss function across iterations. The memorization of previous gradients helps to stabilize updated directions and escape from poor local maxima. We incorporate momentum into the PGD process in Eq. (\ref{eq:pgd process}) as follows:
\begin{equation}
\begin{gathered}
    g_{i} = \mu \cdot g_{i-1} + \frac{\nabla_\delta \mathcal{L}_{\tool}}{\text{mean}(\left\| \nabla_\delta \mathcal{L}_{\tool}\right\|)}, \\
    \delta_{i} = \delta_{i-1} + \gamma \cdot \text{sign}(g_{i}),
\end{gathered}
\label{eq:momentum}
\end{equation}
where $g_{i}$ denotes the accumulated velocity vector at PGD step $i$, and $\mu$ is the decay factor controlling the impact of previous gradients, set to 0.5 by default. In each iteration, the current gradient is normalized using the mean of its absolute values rather than the $L_1$ distance used in the original paper~\cite{dong2018boosting}.

The complete optimization objective of \tool~is:
\begin{equation}
\begin{aligned}
\underset{\| \delta \|_\infty \leq \eta}{arg\,max}\mathcal{L}_{\tool} = \underset{\| \delta \|_\infty \leq \eta}{arg\,max}\lambda_1 \cdot \mathcal{L}_{vae} + \lambda_2 \cdot \mathcal{L}_{feature} \\ + \lambda_3 \cdot \mathcal{L}_{frame} - \lambda_4 \cdot \mathcal{L}_{lpips},
\end{aligned}
\label{eq:dormant loss}
\end{equation}
where $\lambda$s control the weight of each loss term. By default, we set these values as $\lambda_1=10$, $\lambda_2=100$\footnote{For $\mathcal{L}_{feature}$, we set $\lambda_{clip} = 10$ and $\lambda_{ref} = \lambda_2 = 100$, as ReferenceNet has demonstrated the ability to extract more dense and detailed features necessary for appearance alignment compared to the CLIP image encoder~\cite{hu2024animate,xu2024magicanimate}.}, $\lambda_3=1$ and $\lambda_4=10$. 
The detailed optimization process is presented in Alg. \ref{alg:dormant}.

\begin{algorithm}[t]
\caption{\tool}
\begin{algorithmic}[1]
    \Require A human image $x$, PGD iterations $N$, step size $\gamma$, budget $\eta$, decay factor $\mu$; pre-trained models to calculate $\mathcal{L}_{\tool}$: VAE encoder $\mathcal{E}$, CLIP image encoder $\mathcal{C}$, ensemble of ReferenceNets $\mathcal{R}_{1:K}$, and denoising UNet $\epsilon_\theta$ 
    \Ensure The protected image $x_p$
    \State Initialize $g_0 \gets 0$ and $\delta_0 \gets uniform(-\eta,\eta)$
    \For{$i=1$ to $N$}
        \State Sample transformation $\mathcal{T}$
        \State Sample timestep $t$ 
        \State Sample $\tilde{z}_t^{1:F} \sim\mathcal{N}(0,\bm{I})$
        \State Calculate $\mathcal{L}_{\tool}$ by Eq. (\ref{eq:dormant loss})
        \State $g_{i} \gets \mu \cdot g_{i-1} + \frac{\nabla_\delta \mathcal{L}_{\tool}}{\text{mean}(\left\| \nabla_\delta \mathcal{L}_{\tool}\right\|)}$
        \State $\delta_{i} \gets \delta_{i-1} + \gamma \cdot \text{sign}(g_{i})$ 
        \State $\delta_i \gets clip(\delta_i, -\eta, \eta)$
    \EndFor
    \State $x_p \gets x + \delta_N$
    \State \Return $x_p$
\end{algorithmic}
\label{alg:dormant}
\end{algorithm}

\section{Evaluation}
\label{sec:evaluation}

\subsection{Experimental Setup}
\label{subsec:experimental setup}

\begin{figure*}[!t]
\centering
\includegraphics[width=0.977\textwidth]{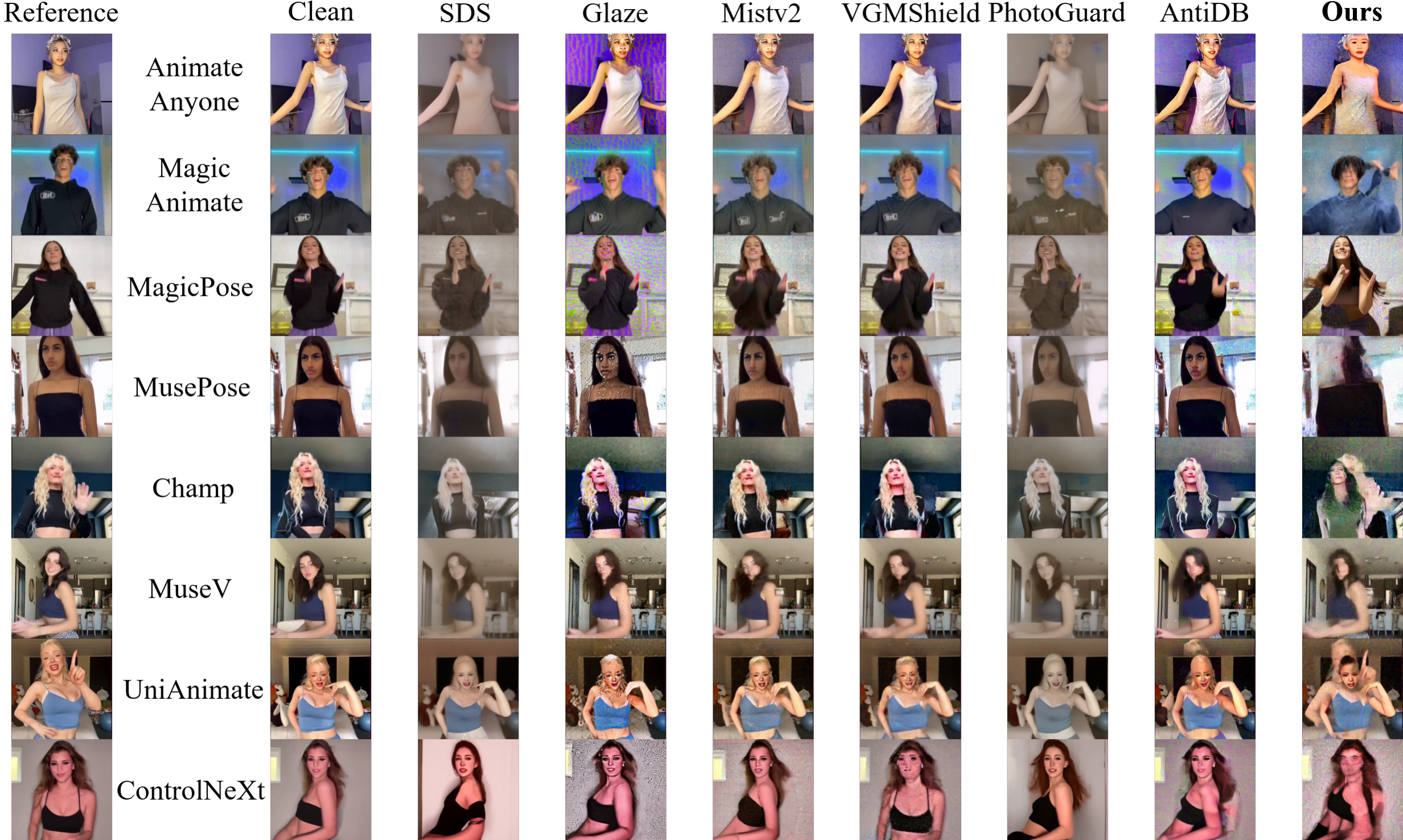}
\caption{Qualitative comparisons with baseline protections against various pose-driven human image animation methods.}
\label{fig:qualitative comparisons with baseline protections}
\end{figure*}

\noindent\textbf{Pose-driven Human Image Animation Methods.} We comprehensively evaluate the effectiveness and transferability of \tool~on 8 cutting-edge and widely-used human image animation methods, including Animate Anyone~\cite{hu2024animate}, MagicAnimate~\cite{xu2024magicanimate}, MagicPose~\cite{chang2024magicpose}, MusePose~\cite{tong2024musepose}, Champ~\cite{zhu2024champ}, MuseV~\cite{xia2024musev}, UniAnimate~\cite{wang2024unianimate}, and ControlNeXt~\cite{peng2024controlnext}. These 8 methods represent SOTA open-source animation techniques within the last two years and cover all major types. Despite advances in controllability and realism, some methods have also been implicated in generating dancing deepfakes on video platforms, raising societal concerns~\cite{pei2024deepfake,liu2024detection,devin2023animate}.

\noindent\textbf{Datasets.} We conduct experiments on 4 datasets: TikTok~\cite{jafarian2021learning}, Champ~\cite{zhu2024champ}, UBC Fashion~\cite{polina2019dwnet}, and TED Talks~\cite{siarohin2021motion}. For the TikTok dataset, we follow the settings used in prior human image animation works~\cite{chang2024magicpose,wang2024disco} and utilize their 10 TikTok-style videos showing different people from the web for evaluation. These videos contain between 248 and 690 frames. Specifically, we use the first frame of each video as the reference image and extract pose sequences from the remaining frames using DWPose~\cite{yang2023effective} or DensePose~\cite{guler2018densepose} to guide the video generation. Additionally, we also sample 10 videos from the Champ training set, and the UBC Fashion and TED Talks test sets for evaluation, resulting in videos with 237-848, 303-355, and 135-259 frames, respectively. All video frames are resized to $512 \times 512$, and we further conduct experiments on other image resolutions, random reference images, and jump-cut pose sequences in Appendix~\ref{sec:protection generality}. 

\noindent\textbf{Baseline Protections.} We compare the protection performance of \tool~with 6 baseline methods defending against text/image-to-image generation (SDS~\cite{xue2024toward}, Glaze 2.1~\cite{shan2023glaze}, Mistv2~\cite{zheng2023improving}, PhotoGuard~\cite{salman2023raising}, and AntiDB~\cite{van2023anti}) and image-to-video generation (VGMShield~\cite{pang2024vgmshield}). For a fair comparison, we set the perturbation budget $\eta=16/255$ and PGD iterations $N=200$ for all protection methods\footnote{Glaze is closed-source software whose perturbation budget cannot be exactly set to $16/255$, we set the intensity to \texttt{High}, the highest setting.}. We evaluate image similarity before and after protections in Appendix~\ref{sec:image similarity}.

\noindent\textbf{Transformations and Purifications.} We evaluate the robustness of \tool~against 5 popular image transformations: JPEG compression, Gaussian blur, Gaussian noise, Median blur, and Bit squeeze. Additionally, we further conduct experiments on 6 advanced purification methods specifically designed to purify the added protective perturbation, including Impress~\cite{cao2024impress}, DiffPure~\cite{nie2022diffusion}, DDSPure~\cite{carlini2023certified}, GrIDPure~\cite{zhao2024can}, Diffshortcut~\cite{liu2024investigating}, and Noisy Upscaling~\cite{honig2025adversarial}.

\noindent\textbf{Metrics.} We evaluate the quality of generated videos using 6 image- and video-wise generative metrics used in prior human image animation methods~\cite{chang2024magicpose,wang2024disco}. For the assessment of image-level quality, we report frame-wise LPIPS~\cite{zhang2018unreasonable}, Fréchet Inception Distance (FID)~\cite{heusel2017gans}, Peak Signal to Noise Ratio (PSNR)~\cite{hore2010image}, and Structural Similarity Index Measure (SSIM)~\cite{wang2004image}; while for video-level evaluation, we concatenate every consecutive 16 frames to form a sample and report FID-VID~\cite{balaji2019conditional} and Fréchet Video Distance (FVD)~\cite{unterthiner2018towards}, respectively.

\noindent\textbf{Platform.} All our experiments are conducted on a server running a 64-bit Ubuntu 22.04.4 system with Intel(R) Xeon(R) Gold 5218R CPU @ 2.10GHz, 512GB memory, and four Nvidia A800 PCIe GPUs with 80GB memory.

We provide detailed introductions to the evaluated animations, protections, transformations, purifications, and image-to-video and image-to-image methods in Appendix~\ref{sec:details of evaluated methods}.

\subsection{Protection Performance}
\label{subsec:protection performance}

\begin{table*}[!t]
\caption{Quantitative comparisons with baseline protections against various pose-driven human image animation methods. $\uparrow$ indicates that a higher value of the metric signifies poorer video quality and thus better protection, while $\downarrow$ indicates the opposite.}
\label{tab:quantitative comparisons with baseline protections}
\centering
\scriptsize
\tabcolsep=0.08cm
\renewcommand\arraystretch{0.914}
\begin{tabular}{>{\centering\arraybackslash}p{2.3cm}>{\centering\arraybackslash}p{1.2cm}>{\centering\arraybackslash}p{1.5cm}>{\centering\arraybackslash}p{1.5cm}>{\centering\arraybackslash}p{1.5cm}>{\centering\arraybackslash}p{1.5cm}>{\centering\arraybackslash}p{1.8cm}>{\centering\arraybackslash}p{1.8cm}>{\centering\arraybackslash}p{1.5cm}>{\centering\arraybackslash}p{1.5cm}}
    \toprule
    {\textbf{Method}} & \textbf{Metric} & \textbf{Clean} & \textbf{SDS~\cite{xue2024toward}} & \textbf{Glaze~\cite{shan2023glaze}} & \textbf{Mistv2~\cite{zheng2023improving}} & \textbf{VGMShield~\cite{pang2024vgmshield}} & \textbf{PhotoGuard~\cite{salman2023raising}} & \textbf{AntiDB~\cite{van2023anti}} & \textbf{\tool} \\ 
    \midrule
    \multirow{6}{*}{\textbf{Animate Anyone~\cite{hu2024animate}}} & 
    \textbf{FID-VID$\uparrow$} &41.61 & 108.40 & 109.98 & 59.49 & 53.95 & 112.98 & 84.75 & \textbf{162.87} \\
    & \textbf{FVD$\uparrow$} & 362.75 & 859.75 & 1301.96 & 678.56 & 625.83 & 994.84 & 1055.58 & \textbf{1364.00} \\
    & \textbf{LPIPS$\uparrow$} & 0.282 & 0.556 & 0.582 & 0.554 & 0.584 & 0.565 & 0.542 & \textbf{0.639} \\
    & \textbf{FID$\uparrow$} & 65.00 & 140.69 & 172.04 & 142.04 & 142.07 & 143.41 & 181.52 & \textbf{245.06} \\
    & \textbf{PSNR$\downarrow$} &17.76 & 16.30 & 15.23 & 17.38 & 16.75 & 15.94 & 15.76 & \textbf{11.82} \\
    & \textbf{SSIM$\downarrow$} & 0.741 & 0.684 & 0.446 & 0.553 & 0.510 & 0.682 & 0.441 & \textbf{0.374} \\
    \midrule
    \multirow{6}{*}{\textbf{MagicAnimate~\cite{xu2024magicanimate}}} & 
    \textbf{FID-VID$\uparrow$} &37.04 & 148.59 & 92.20 & 66.49 & 56.00 & 145.01 & 60.61 & \textbf{174.93} \\
    & \textbf{FVD$\uparrow$} & 374.99 & 1051.55 & 1002.92 & 616.33 & 409.64 & 1081.05 & 620.48 & \textbf{1174.52} \\
    & \textbf{LPIPS$\uparrow$} & 0.268 & 0.409 & 0.531 & 0.499 & 0.506 & 0.421 & 0.502 & \textbf{0.604} \\
    & \textbf{FID$\uparrow$} & 68.86 & 131.53 & 132.25 & 119.88 & 111.03 & 127.21 & 115.07 & \textbf{235.23} \\
    & \textbf{PSNR$\downarrow$} &18.29 & 15.03 & 15.74 & 17.33 & 17.72 & 14.78 & 17.18 & \textbf{13.89} \\
    & \textbf{SSIM$\downarrow$} & 0.758 & 0.676 & 0.493 & 0.595 & 0.600 & 0.675 & 0.535 & \textbf{0.369} \\
    \midrule
    \multirow{6}{*}{\textbf{MagicPose~\cite{chang2024magicpose}}} & 
    \textbf{FID-VID$\uparrow$} &67.73 & 98.76 & 120.89 & 107.22 & 97.04 & \textbf{134.99} & 98.76 & 133.06 \\
    & \textbf{FVD$\uparrow$} & 431.39 & 902.56 & 1058.04 & 899.49 & 764.42 & 1013.40 & 902.56 & \textbf{1413.94} \\
    & \textbf{LPIPS$\uparrow$} & 0.291 & 0.523 & 0.525 & 0.505 & 0.517 & 0.434 & 0.523 & \textbf{0.575} \\
    & \textbf{FID$\uparrow$} & 55.06 & 132.95 & 130.18 & 119.35 & 105.56 & 116.43 & 132.95 & \textbf{176.29} \\
    & \textbf{PSNR$\downarrow$} &17.24 & 16.06 & 15.67 & 16.21 & 16.46 & 15.58 & 16.06 & \textbf{13.77} \\
    & \textbf{SSIM$\downarrow$} & 0.745 & 0.506 & 0.510 & 0.534 & 0.550 & 0.681 & 0.506 & \textbf{0.432} \\
    \midrule
    \multirow{6}{*}{\textbf{MusePose~\cite{tong2024musepose}}} & 
    \textbf{FID-VID$\uparrow$} &30.17 & 73.77 & 107.58 & 52.87 & 47.93 & 92.29 & 73.77 & \textbf{182.47} \\
    & \textbf{FVD$\uparrow$} & 343.91 & 840.79 & 1208.92 & 556.01 & 441.20 & 800.16 & 840.79 & \textbf{1265.38} \\
    & \textbf{LPIPS$\uparrow$} & 0.284 & 0.524 & 0.574 & 0.532 & 0.560 & 0.462 & 0.524 & \textbf{0.642} \\
    & \textbf{FID$\uparrow$} & 66.08 & 147.22 & 163.89 & 126.60 & 128.12 & 126.95 & 147.22 & \textbf{280.57} \\
    & \textbf{PSNR$\downarrow$} &17.77 & 16.08 & 15.12 & 17.46 & 17.42 & 16.67 & 16.08 & \textbf{11.50} \\
    & \textbf{SSIM$\downarrow$} & 0.745 & 0.481 & 0.461 & 0.601 & 0.572 & 0.712 & 0.481 & \textbf{0.381} \\
    \midrule
    \multirow{6}{*}{\textbf{Champ~\cite{zhu2024champ}}} & 
    \textbf{FID-VID$\uparrow$} &85.19 & 144.27 & 148.12 & 107.79 & 97.79 & 152.46 & 111.93 & \textbf{158.70} \\
    & \textbf{FVD$\uparrow$} & 498.01 & 1017.24 & \textbf{1291.57} & 806.41 & 779.76 & 1119.71 & 1062.11 & 1234.56 \\
    & \textbf{LPIPS$\uparrow$} & 0.394 & 0.498 & 0.632 & 0.605 & 0.625 & 0.510 & 0.600 & \textbf{0.655} \\
    & \textbf{FID$\uparrow$} & 71.66 & 126.57 & 153.60 & 123.60 & 130.77 & 130.89 & 167.54 & \textbf{233.61} \\
    & \textbf{PSNR$\downarrow$} &13.29 & 13.68 & 11.79 & 13.03 & 12.62 & 13.63 & 12.00 & \textbf{11.57} \\
    & \textbf{SSIM$\downarrow$} & 0.642 & 0.625 & 0.391 & 0.496 & 0.470 & 0.621 & 0.415 & \textbf{0.349} \\
    \midrule
    \multirow{6}{*}{\textbf{MuseV~\cite{xia2024musev}}} & 
    \textbf{FID-VID$\uparrow$} &60.75 & 128.61 & 98.63 & 74.06 & 89.16 & 119.82 & 89.46 & \textbf{151.46} \\
    & \textbf{FVD$\uparrow$} & 430.96 & 820.45 & 987.90 & 601.51 & 603.88 & 852.04 & 622.80 & \textbf{1078.03} \\
    & \textbf{LPIPS$\uparrow$} & 0.310 & 0.563 & 0.543 & 0.533 & 0.539 & 0.512 & 0.407 & \textbf{0.607} \\
    & \textbf{FID$\uparrow$} & 82.19 & 127.12 & 139.48 & 115.29 & 117.65 & 115.89 & 136.73 & \textbf{240.82} \\
    & \textbf{PSNR$\downarrow$} &16.11 & 15.79 & 15.22 & 16.32 & 15.78 & 15.98 & 15.36 & \textbf{14.48} \\
    & \textbf{SSIM$\downarrow$} & 0.721 & 0.665 & 0.563 & 0.649 & 0.644 & 0.681 & 0.675 & \textbf{0.526} \\
    \midrule
    \multirow{6}{*}{\textbf{UniAnimate~\cite{wang2024unianimate}}} & 
    \textbf{FID-VID$\uparrow$} &26.03 & 86.59 & 105.43 & 49.75 & 43.15 & 81.76 & 56.77 & \textbf{113.73} \\
    & \textbf{FVD$\uparrow$} & 277.02 & 704.07 & \textbf{1134.13} & 572.21 & 490.22 & 667.84 & 708.05 & 1031.22 \\
    & \textbf{LPIPS$\uparrow$} & 0.253 & 0.521 & 0.579 & 0.566 & 0.567 & 0.488 & 0.505 & \textbf{0.580} \\
    & \textbf{FID$\uparrow$} & 52.63 & 131.51 & 153.04 & 115.08 & 119.63 & 123.60 & 167.66 & \textbf{250.30} \\
    & \textbf{PSNR$\downarrow$} &18.76 & 15.76 & 15.58 & 17.91 & 17.98 & 16.09 & 17.40 & \textbf{15.57} \\
    & \textbf{SSIM$\downarrow$} & 0.762 & 0.680 & 0.474 & 0.566 & 0.587 & 0.709 & 0.545 & \textbf{0.470} \\
    \midrule
    \multirow{6}{*}{\textbf{ControlNeXt~\cite{peng2024controlnext}}} & 
    \textbf{FID-VID$\uparrow$} &35.73 & 105.07 & 96.45 & 72.94 & 84.25 & 80.65 & 79.35 & \textbf{110.37} \\
    & \textbf{FVD$\uparrow$} & 337.31 & 1001.21 & 1134.70 & 764.49 & 862.51 & 775.46 & 839.82 & \textbf{1141.36} \\
    & \textbf{LPIPS$\uparrow$} & 0.301 & 0.478 & 0.501 & 0.494 & \textbf{0.541} & 0.428 & 0.492 & 0.538 \\
    & \textbf{FID$\uparrow$} & 62.87 & 130.81 & 126.56 & 106.19 & 140.87 & 112.98 & 138.89 & \textbf{163.06} \\
    & \textbf{PSNR$\downarrow$} &16.88 & \textbf{11.95} & 13.88 & 14.44 & 13.92 & 13.55 & 14.85 & 13.19 \\
    & \textbf{SSIM$\downarrow$} & 0.734 & 0.529 & 0.505 & 0.547 & 0.485 & 0.617 & 0.532 & \textbf{0.441} \\
    \bottomrule	
\end{tabular}
\end{table*}

\noindent\textbf{Overall Results.} We evaluate the protection effectiveness and transferability of \tool~against eight pose-driven human image animation methods on the TikTok dataset, with the quantitative results shown in Table~\ref{tab:quantitative comparisons with baseline protections}. Evaluation across six metrics demonstrates significant degradation in the quality of generated videos due to the protective effect of \tool~on all eight animation methods. Specifically, the average FID-VID, FVD, LPIPS, and FID increase from 48.03, 382.04, 0.30, and 65.54 to 148.45, 1212.88, 0.61 and 228.12, and the average PSNR and SSIM decrease from 17.01 and 0.731 to 13.22 and 0.418, respectively. We show non-cherry-picked generated video frames in Figure~\ref{fig:qualitative comparisons with baseline protections}, which display mismatched identities, distorted backgrounds, and visual artifacts. More visualization results for videos of \tool~against various animation techniques can be found in Figure~\ref{fig:more visualizations of method} in Appendix~\ref{sec:more visualized results of videos}, which exhibit inconsistencies among frames. These results highlight the effectiveness of \tool~in preventing unauthorized image usage in human video generation, thereby safeguarding individuals' rights to portrait and privacy.

We also compare \tool's protection performance with six baseline methods that defend against unauthorized image and video generation. Quantitative results in Table~\ref{tab:quantitative comparisons with baseline protections} and qualitative results in Figure~\ref{fig:qualitative comparisons with baseline protections} demonstrate the superior protection performance of \tool, showing the greatest effectiveness in degrading the quality of generated videos. While Dormant exhibits the strongest transferability and performs effectively across all 8 animation techniques, baseline protections show limited transferability and are only effective for certain animation methods, yielding some better metrics with different protective effects. Specifically, the concurrent work VGMShield achieves slightly higher LPIPS on the SVD-based method ControlNeXt but lacks transferability to other animation methods. SDS tends to lighten colors and smooth out details, resulting in better PSNR on ControlNeXt. PhotoGuard induces noticeable inconsistencies between adjacent frames, showing slightly higher FID-VID on MagicPose. Glaze produces visual artifacts and intensifies colors in generated videos, leading to better FVD on Champ and UniAnimate. Moreover, although these baseline protections may perform better on some metrics for certain animation methods, \tool~still 1) shows a clear advantage in other metrics for the corresponding animation method; and 2) outperforms them across all other animation methods. 

\begin{figure}[t]
   \centering
   \includegraphics[width=0.47\textwidth]{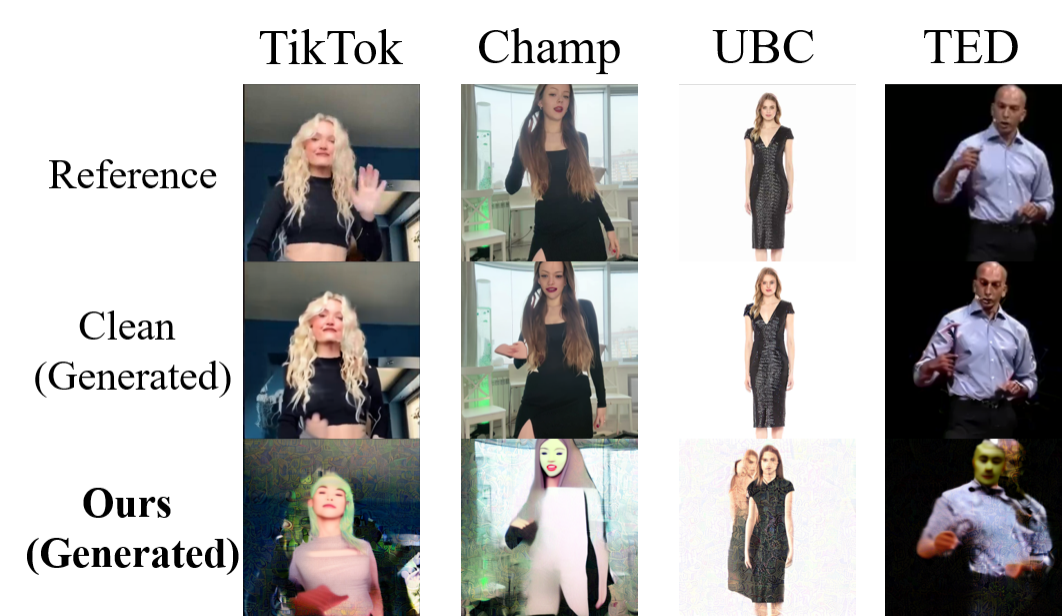}
    \caption{Qualitative results on various datasets.}
    \label{fig:qualitative results on various datasets}
\end{figure}

\noindent\textbf{Results on Various Datasets.} Besides the TikTok dataset, we also evaluate the protection performance of \tool~on three other datasets using Animate Anyone, including human dance generation on the Champ dataset, fashion video synthesis on the UBC Fashion dataset, and speech video generation on the TED Talks dataset. Quantitative results in Table~\ref{tab:protection performance on various datasets} demonstrate the remarkable performance of \tool~in defending against pose-driven human image animation, with the average FID-VID, FVD, LPIPS and FID increasing from 53.78, 406.03, 0.22 and 60.81 to 166.28, 1288.20, 0.57 and 260.53, and the average PSNR and SSIM decreasing from 19.44 and 0.745 to 14.69 and 0.486, respectively. The poor-quality video frames shown in Figure~\ref{fig:qualitative results on various datasets} also validate the protection effectiveness of \tool. More visualization results for videos can be found in Figure~\ref{fig:more visualizations of dataset} in Appendix~\ref{sec:more visualized results of videos}.

\subsection{Human and GPT-4o Studies}
\label{subsec:human and gpt-4o studies}

\noindent\textbf{Human Study.} To bridge the gap between quantitative metrics and human perception, as well as study whether \tool~outperforms baseline methods in protecting privacy and portrait rights from the perspective of real-world audiences, we conduct a survey study. Our questionnaire consists of two parts: 1) demographic data collection, including five questions on age, gender, education, expertise, and familiarity; and 2) ten questions asking participants to select the video that best protects portrait and privacy rights compared to the reference, each question containing a reference video and five videos generated from images protected by \tool~and four baseline methods (SDS, Glaze, VGMShield, and PhotoGuard). The videos used are the 10 videos in the TikTok test set, derived from experiments conducted using Animate Anyone, with the corresponding metrics provided in Table~\ref{tab:quantitative comparisons with baseline protections}. The detailed questionnaire can be found in Appendix~\ref{subsec:questionnaire in human study}. 

\begin{table}[t]
\caption{Protection performance on various datasets.}
\label{tab:protection performance on various datasets}
\centering
\scriptsize
\tabcolsep=0.078cm
\renewcommand\arraystretch{1}
\begin{tabular}{cccccccc}
    \toprule
    \multicolumn{2}{c}{\textbf{Dataset}} &\textbf{FID-VID$\uparrow$} &\textbf{FVD$\uparrow$} &\textbf{PSNR$\downarrow$} &\textbf{SSIM$\downarrow$} &\textbf{LPIPS$\uparrow$} &\textbf{FID$\uparrow$}\\
    \midrule
    \multirow{2}{*}{\textbf{TikTok~\cite{jafarian2021learning}}} & \textbf{Clean} & 41.61 & 362.75 & 17.76 & 0.741 & 0.282 & 65.00 \\
    & \textbf{Protect} & 162.87 & 1364.00 & 11.82 & 0.374 & 0.639 & 245.06 \\
    \midrule
    \multirow{2}{*}{\textbf{Champ~\cite{zhu2024champ}}} & \textbf{Clean} & 45.92 & 379.64 & 18.04 & 0.661 & 0.286 & 56.72 \\
    & \textbf{Protect} & 156.03 & 1465.48 & 13.72 & 0.397 & 0.597 & 252.03 \\
    \midrule
    \multirow{2}{*}{\makecell{\textbf{UBC} \\ \textbf{Fashion~\cite{polina2019dwnet}}}} & \textbf{Clean} & 38.04 & 329.28 & 20.39 & 0.876 & 0.086 & 33.74 \\
    & \textbf{Protect} & 78.67 & 780.96 & 18.38 & 0.692 & 0.432 & 164.99 \\
    \midrule
    \multirow{2}{*}{\textbf{TED Talks~\cite{siarohin2021motion}}} & \textbf{Clean} & 89.54 & 552.45 & 21.56 & 0.701 & 0.233 & 87.78 \\
    & \textbf{Protect} & 267.54 & 1542.37 & 14.82 & 0.479 & 0.627 & 380.04 \\
    \bottomrule	
\end{tabular}
\end{table}

We distributed the questionnaire online and recruited participants via social media, offering lottery-based compensation upon completion. A total of 82 valid responses were collected. The participants' ages primarily ranged from 18 to 54, with the majority (63.41\%) falling between 25 and 34. Among the participants, 57.32\% were male and 40.24\% were female; 80.48\% held or were pursuing a bachelor's degree or higher; 39.02\% had a background in computer-related fields; and 70.73\% had at least heard of image-to-video generation or were more familiar with it. As shown in Table~\ref{tab:results of human and GPT-4o studies}, \tool~achieved an overall pick rate of 74.27\% (609/820), significantly outperforming baseline protections. Detailed results for each question are shown in Figure~\hyperref[fig:results of human study]{15(a)} in Appendix~\ref{subsec:detailed results}. Notably, 67 participants selected \tool~for at least five questions, including 29 with a computer-related background (32 in total). 37 participants selected \tool~for all ten questions, while only 8 participants did not choose \tool~for any of the ten questions, with SDS being their preferred method, yielding a pick rate of 72.50\% (58/80). Of these eight participants, only one had heard of image-to-video generation, and the others were less familiar with it.

\begin{table}[htbp]
\caption{Results of human and GPT-4o studies.}
\label{tab:results of human and GPT-4o studies}
\centering
\scriptsize
\tabcolsep=0.08cm
\renewcommand\arraystretch{1}
\begin{tabular}{>{\centering\arraybackslash}p{2cm}>{\centering\arraybackslash}p{0.8cm}>{\centering\arraybackslash}p{0.8cm}>{\centering\arraybackslash}p{1.3cm}>{\centering\arraybackslash}p{1.3cm}>{\centering\arraybackslash}p{1.3cm}}
    \toprule
    {\textbf{Metric}} &\textbf{SDS} & \textbf{Glaze} & \textbf{VGMShield} & \textbf{PhotoGuard} & \textbf{\tool}\\
    \midrule
    \textbf{Human Pick Rate$\uparrow$} & 10.12\% & 9.51\% & 4.27\% & 1.83\% & \textbf{74.27\%}\\
    \textbf{GPT-4o Rank$\downarrow$} & 4.0  & 2.0 & 2.7 & 4.8 & \textbf{1.5}\\
    \bottomrule	
\end{tabular}
\end{table}

\begin{figure*}[ht]
  \centering
  \begin{subfigure}[b]{0.196\linewidth}
        \centering
        \includegraphics[width=\textwidth]{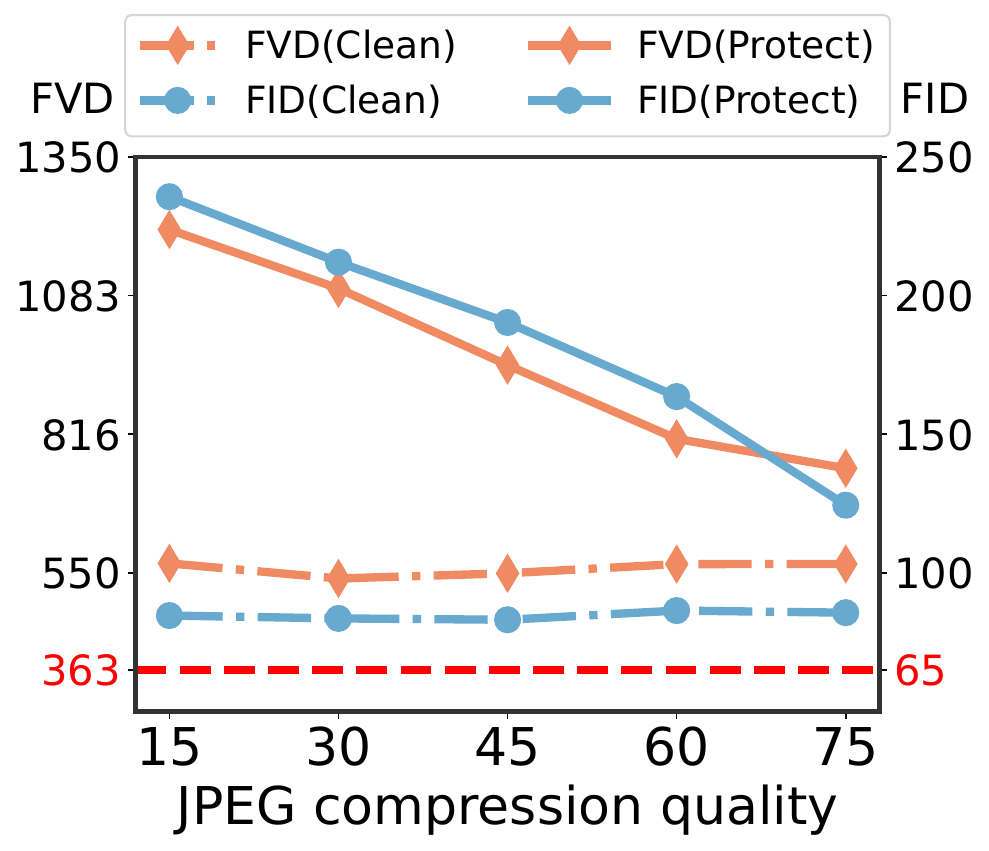}
        \caption{JPEG compression}
        \label{fig:jpeg compression}
  \end{subfigure}
  \hfill
  \begin{subfigure}[b]{0.196\linewidth}
        \centering
        \includegraphics[width=\textwidth]{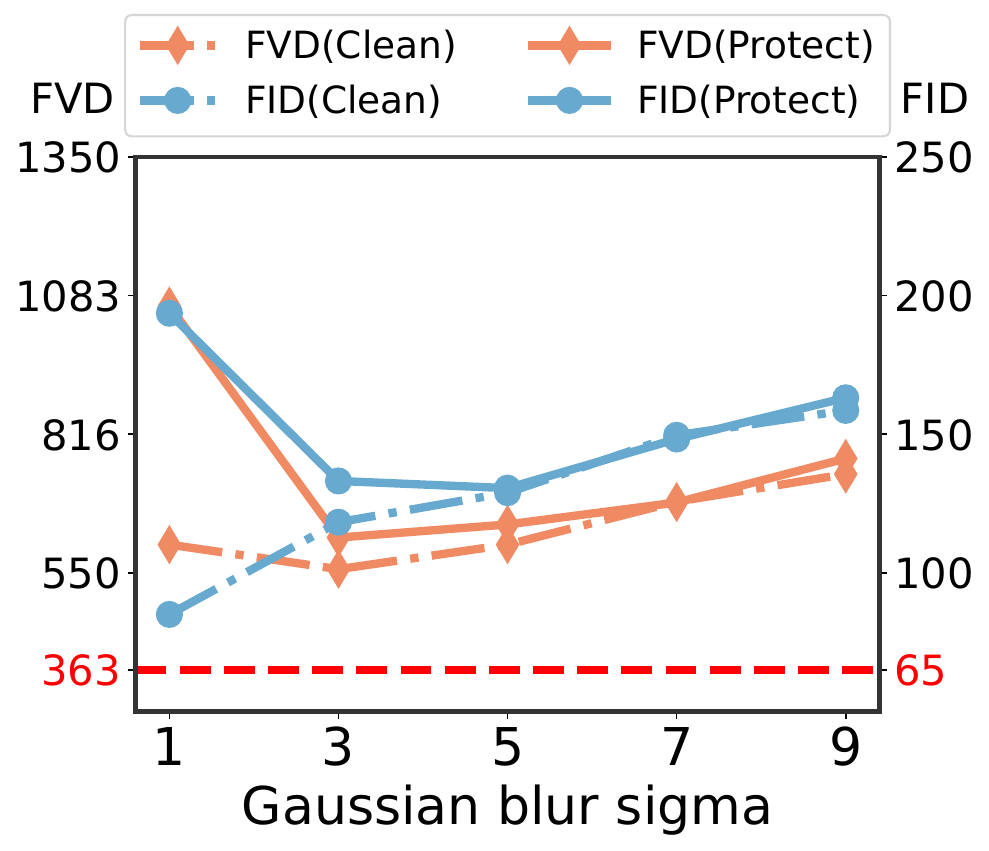}
        \caption{Gaussian blur}
        \label{fig:gaussian blur}
  \end{subfigure}
  \hfill
  \begin{subfigure}[b]{0.196\linewidth}
        \centering
        \includegraphics[width=\textwidth]{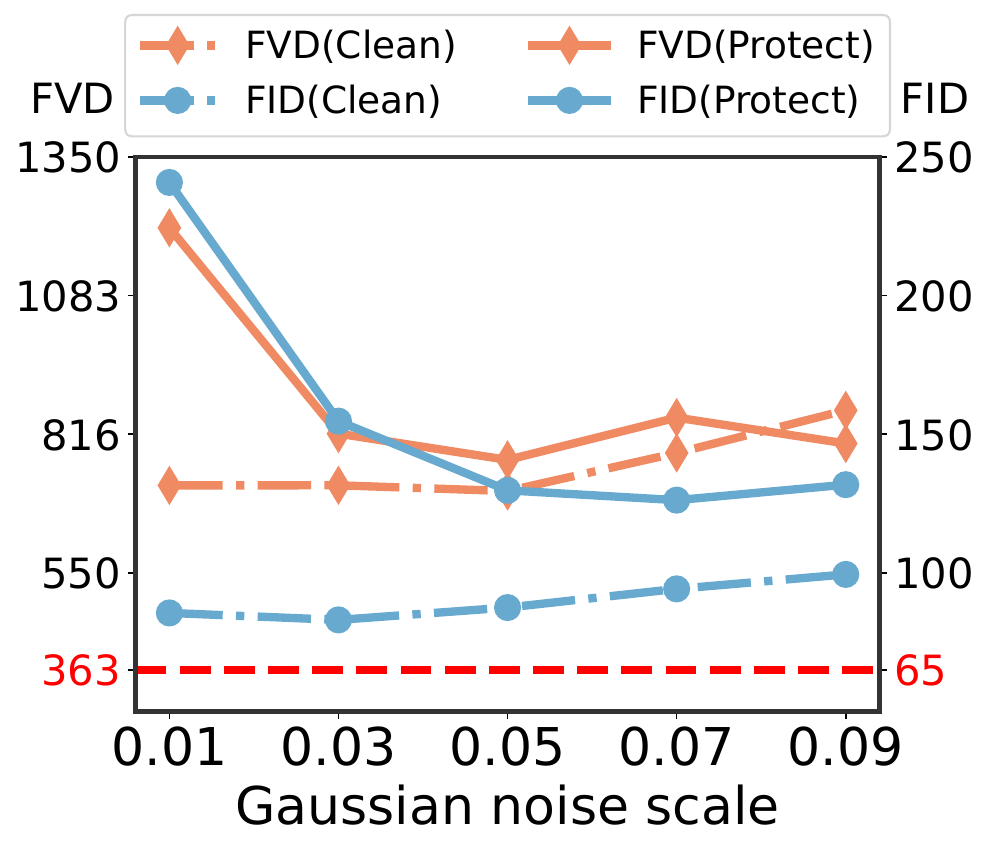}
        \caption{Gaussian noise}
        \label{fig:gaussian noise}
  \end{subfigure}
  \hfill
  \begin{subfigure}[b]{0.196\linewidth}
        \centering
        \includegraphics[width=\textwidth]{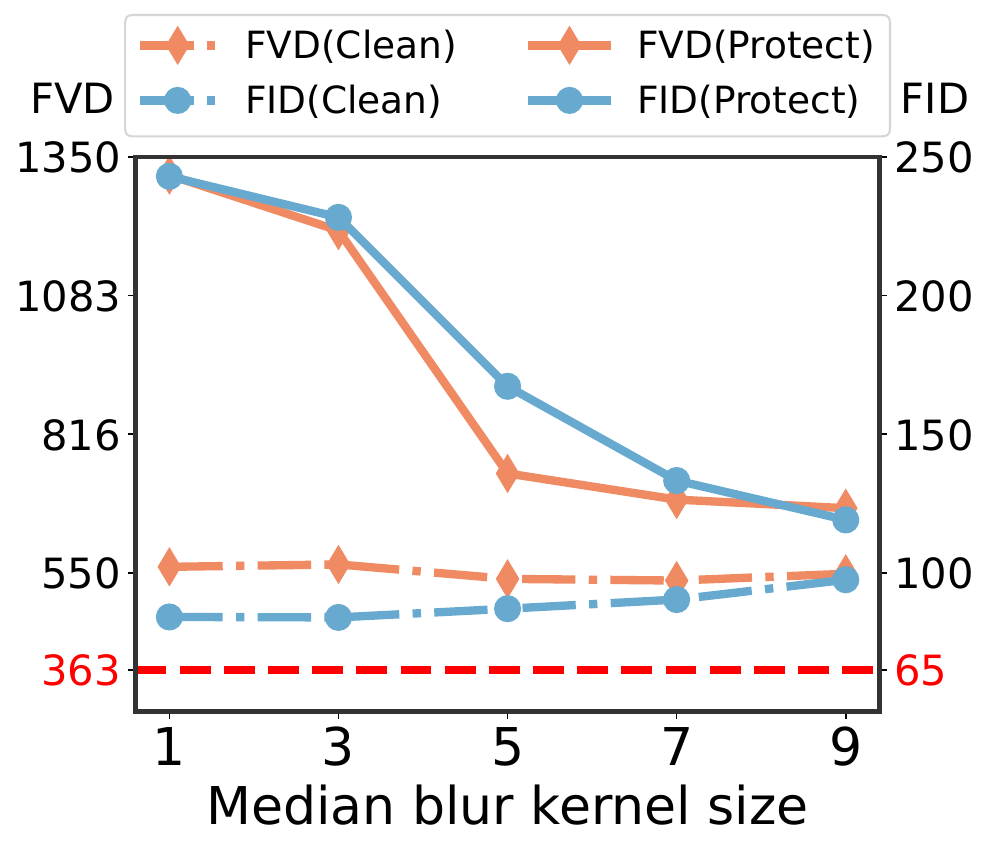}
        \caption{Median blur}
        \label{fig:median blur}
  \end{subfigure}
  \hfill
  \begin{subfigure}[b]{0.196\linewidth}
        \centering
        \includegraphics[width=\textwidth]{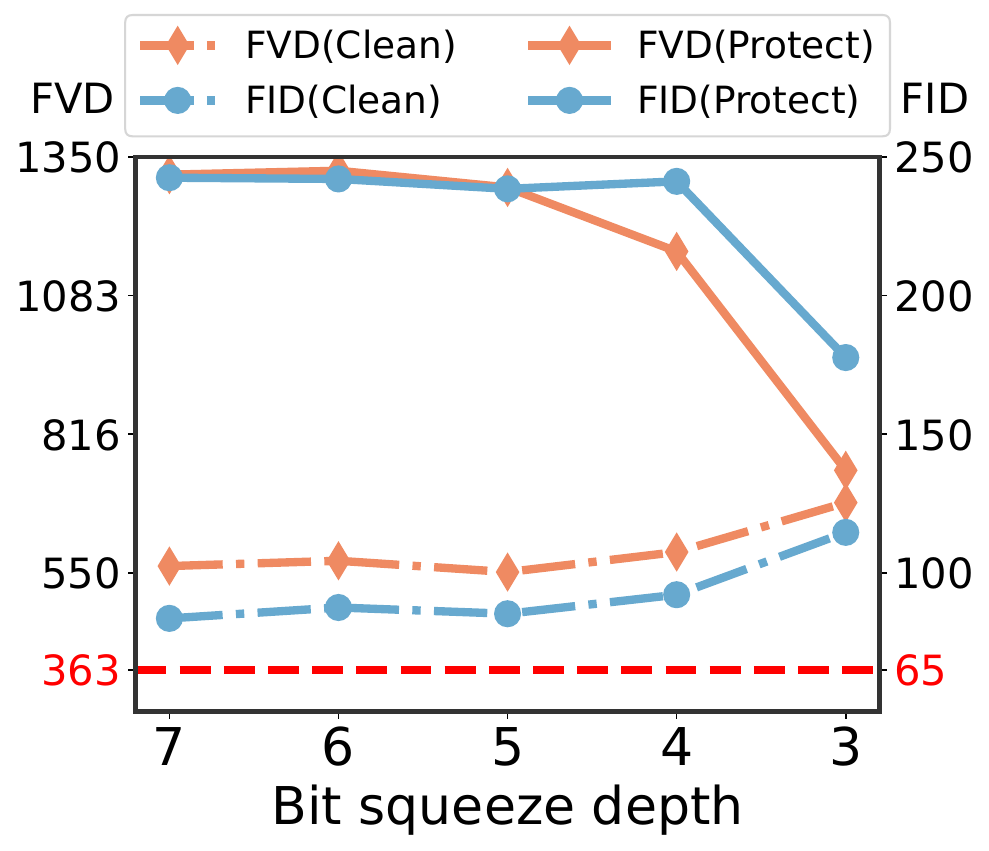}
        \caption{Bit squeeze}
        \label{fig:bit squeeze}
  \end{subfigure}
  \caption{Protection robustness of \tool~against various transformations under different parameter settings.}
  \label{fig:protection robustness against various transformations}
\end{figure*}

\noindent\textbf{GPT-4o Study.} As LLM-as-a-Judge is now widely adopted in many complex tasks~\cite{gu2024survey}, we also use GPT-4o~\cite{openai2024gpt-4o} to simulate the above human study. GPT-4o is a multimodal large language model capable of understanding image semantics and perceiving pixel-level differences. We input GPT-4o with a reference and five generated frame sequences, asking it to rank them based on their effectiveness in protecting portrait and privacy rights, focusing mainly on factors such as appearance mismatches and background changes that differentiate the generated frames from the reference. GPT-4o is instructed to analyze step by step and output both the ranking and its reasoning. Our prompt is detailed in Appendix~\ref{subsec:prompt in gpt-4o study}. We further manually verify all outputs from GPT-4o to ensure there are no hallucinations. As shown in Table~\ref{tab:results of human and GPT-4o studies}, \tool~achieves the highest average ranking of 1.5, and is consistently ranked first or second across all samples (see Figure~\hyperref[fig:results of GPT-4o study]{15(b)} in Appendix~\ref{subsec:detailed results}). While GPT-4o ranks \tool~first, it notes that this method ``makes the individual unrecognizable and drastically alters the background''. Glaze is preferred by GPT-4o when \tool~is ranked second, as it ``incorporates more prominent and deliberate texture patterns''.

\begin{figure}[t]
   \centering
   \includegraphics[width=0.47\textwidth]{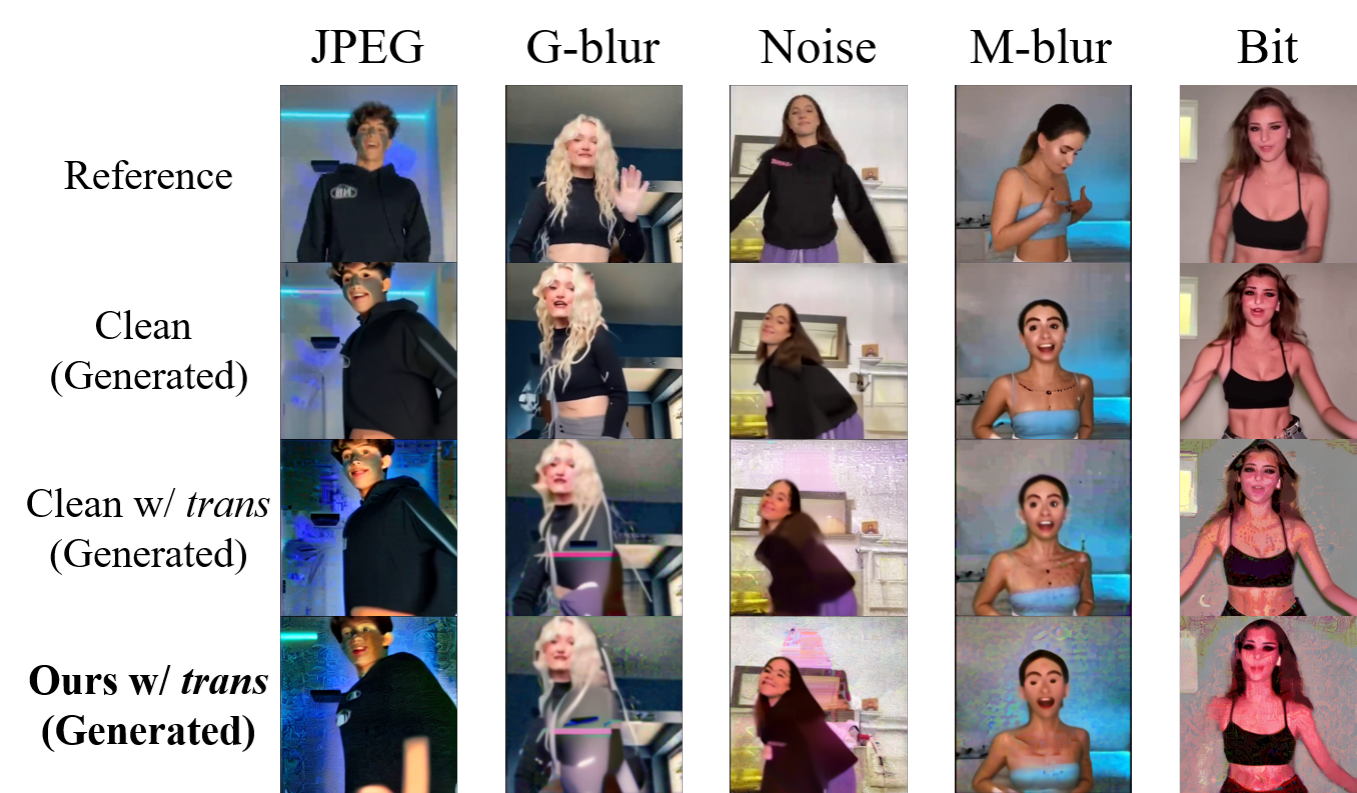}
    \caption{Qualitative results of transformations that most degrade the protection, with JPEG compression (quality=75), Gaussian blur (sigma=3), Gaussian noise (scale=0.05), Median blur (kernel size=9) and Bit squeeze (depth=3).}
    \label{fig:qualitative results against transformations}
\end{figure}

\subsection{Protection Robustness}
\label{subsec:protection robustness}

\noindent\textbf{Robustness against Transformations.} We evaluate five transformations commonly used in prior studies~\cite{zhao2024can,honig2025adversarial,ding2019advertorch}, including three transformations applied in EoT (JPEG compression, Gaussian blur, and Gaussian noise) and two unseen transformations (Median blur and Bit squeeze). We apply them with different parameter settings to both the original and protected images to investigate their impact on the quality of generated videos and the effectiveness of protection. As shown in Figure~\ref{fig:protection robustness against various transformations} (with the red dotted line representing the results of videos generated using the original image without transformations), the results indicate that: 1) transformations can only reduce the protection effect to some extent but cannot completely remove the impact of the protective perturbation; 2) transformations also affect the quality of generated videos. In particular, increasing the transformation parameters to minimize the protection effect would also noticeably degrade the quality of the generated videos. Figure~\ref{fig:qualitative results against transformations} presents the qualitative results when various transformations most degrade the protection effect. Even in these cases, the generated videos still exhibit low quality due to residual protective perturbation and the impact of excessive transformations, displaying issues including distorted backgrounds, mismatched faces, and visual artifacts. These results validate the robustness of \tool~against various transformations.

\begin{figure}[t]
   \centering
   \includegraphics[width=0.47\textwidth]{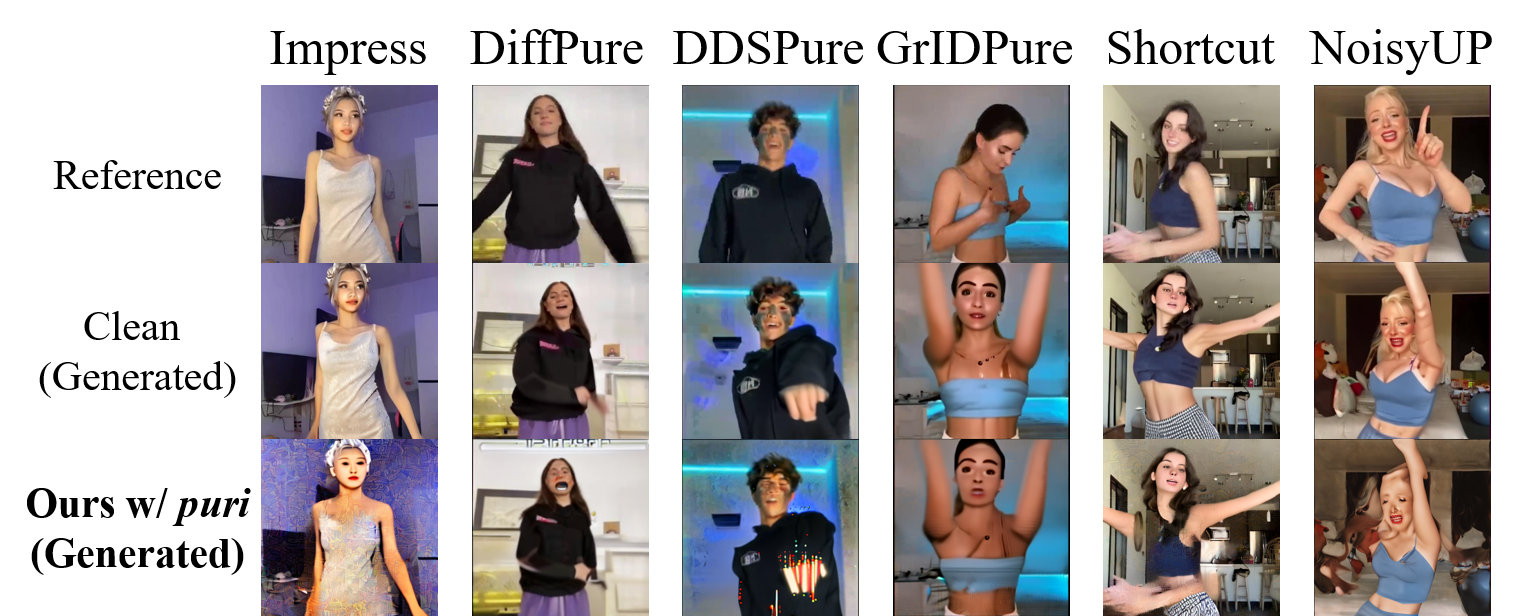}
    \caption{Qualitative results against various purifications.}
    \label{fig:qualitative results against purifications}
\end{figure}

\begin{table}[t]
\caption{Quantitative results against various purifications.}
\label{tab:quantitative results against various purifications}
\centering
\scriptsize
\tabcolsep=0.08cm
\renewcommand\arraystretch{1}
\begin{tabular}{ccccccc}
    \toprule
    {\textbf{Method}} &\textbf{FID-VID$\uparrow$} &\textbf{FVD$\uparrow$} &\textbf{PSNR$\downarrow$} &\textbf{SSIM$\downarrow$} &\textbf{LPIPS$\uparrow$} &\textbf{FID$\uparrow$}\\
    \midrule
    \textbf{Impress~\cite{cao2024impress}} & 171.49 & 1386.19 & 13.31 & 0.341 & 0.628 & 259.26 \\
    \textbf{DiffPure~\cite{nie2022diffusion}} & 54.43 & 490.63 & 17.57 & 0.712 & 0.337 & 94.23 \\
    \textbf{DDSPure~\cite{carlini2023certified}} & 71.38 & 762.01 & 17.14 & 0.687 & 0.398 & 126.93 \\
    \textbf{GrIDPure~\cite{zhao2024can}} & 62.69 & 465.87 & 16.78 & 0.670 & 0.338 & 101.40 \\
    \textbf{Diffshortcut~\cite{liu2024investigating}} & 73.48 & 605.54 & 16.11 & 0.545 & 0.514 & 152.60 \\
    \textbf{Noisy Upscaling~\cite{honig2025adversarial}} & 78.10 & 673.03 & 17.02 & 0.680 & 0.414 & 144.24 \\
    \midrule
    \textbf{Clean} & \textbf{41.61} & \textbf{362.75} & \textbf{17.76} & \textbf{0.741} & \textbf{0.282} & \textbf{65.00} \\
    \bottomrule	
\end{tabular}
\end{table}

\noindent\textbf{Robustness against Purifications.} We evaluate the robustness of \tool~against six purification methods, and the results are shown in Table~\ref{tab:quantitative results against various purifications}. While the protective effect of \tool~is somewhat reduced by these purifications, there still remains a noticeable quality gap between the resulting videos and those generated from the original images. Qualitative results are presented in Figure~\ref{fig:qualitative results against purifications}. Impress shows almost no purifying effect. DiffPure and DDSPure would introduce external strange patterns and light streaks into the generated videos. GrIDPure and Noisy Upscaling fail to restore detailed facial features. In contrast, while Diffshortcut utilizes CodeFormer~\cite{zhou2022towards} to restore human faces (which are vivid but differ from the reference), there are still noticeable artifacts in other regions. As a result, videos generated using protected images after purification cannot be used directly for malicious purposes due to their low quality, and the individuals' privacy and portrait rights are still effectively safeguarded.

\begin{figure}[t]
   \centering
   \includegraphics[width=0.435\textwidth]{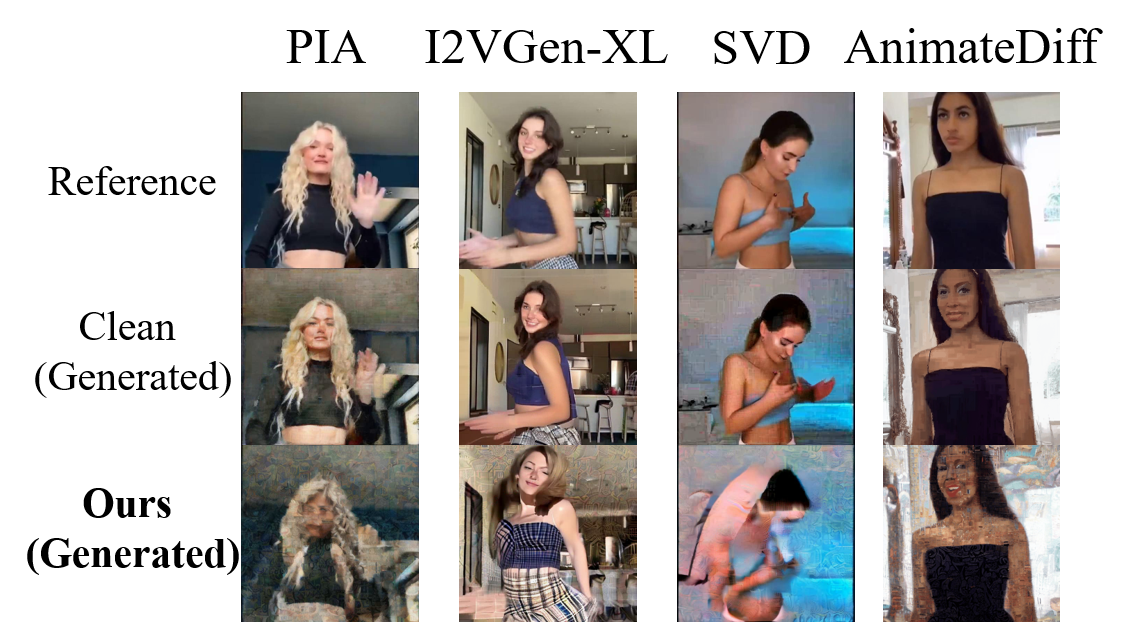}
    \caption{Qualitative results on image-to-video methods.}
    \label{fig:qualitative results on image-to-video methods}
\end{figure}

\begin{figure}[t]
   \centering
   \includegraphics[width=0.435\textwidth]{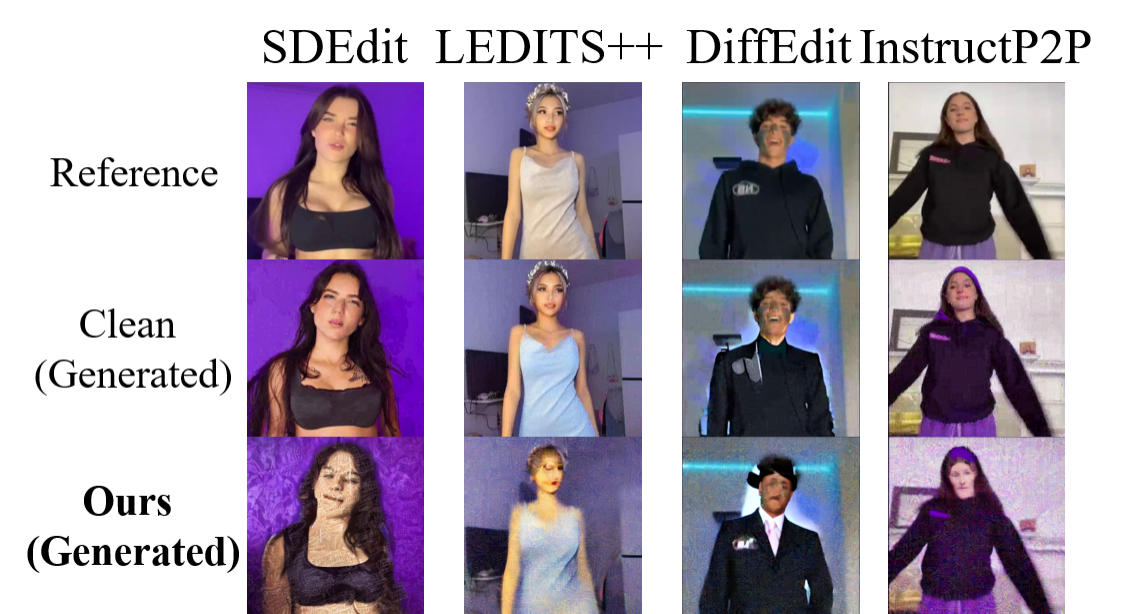}
    \caption{Qualitative results on image-to-image methods.}
    \label{fig:qualitative results on image-to-image methods}
\end{figure}

\subsection{Protection Transferability}
\label{subsec:protection transferability}

\noindent\textbf{Results on Image-to-Video/Image.} Besides human image animation methods that use pose sequences to guide video generation from the reference image, we conduct additional experiments on four other types of image-to-video methods that directly generate videos from the given image without external guidance (SVD) and those that use text prompts for guidance (PIA, I2VGen-XL, and AnimateDiff). Moreover, we also consider the scenario of image manipulation and evaluate the performance of \tool~using four image-to-image methods that use text prompts as conditions for image editing. We utilize BLIP-2~\cite{li2023blip} to generate captions corresponding to the reference images, resulting in prompts such as ``A woman in a white dress''. To guide video generation, we add the suffix ``\textit{is dancing}'' to these captions. For image-to-image methods, we use the original captions for SDEdit and InstructPix2Pix, while for LEDITS++ and DiffEdit, we generate variant prompts using GPT-4o, such as ``A woman in a \textit{blue} dress''. Since FID-VID and FVD cannot be calculated here, we employ two other metrics, CLIP-I and DINO, which compute the average pairwise cosine similarity between CLIP image embeddings~\cite{radford2021learning} and ViT-S/16 DINO embeddings~\cite{caron2021emerging} of generated and reference images, respectively.

\begin{figure}[t]
    \centering
    \includegraphics[width=0.47\textwidth]{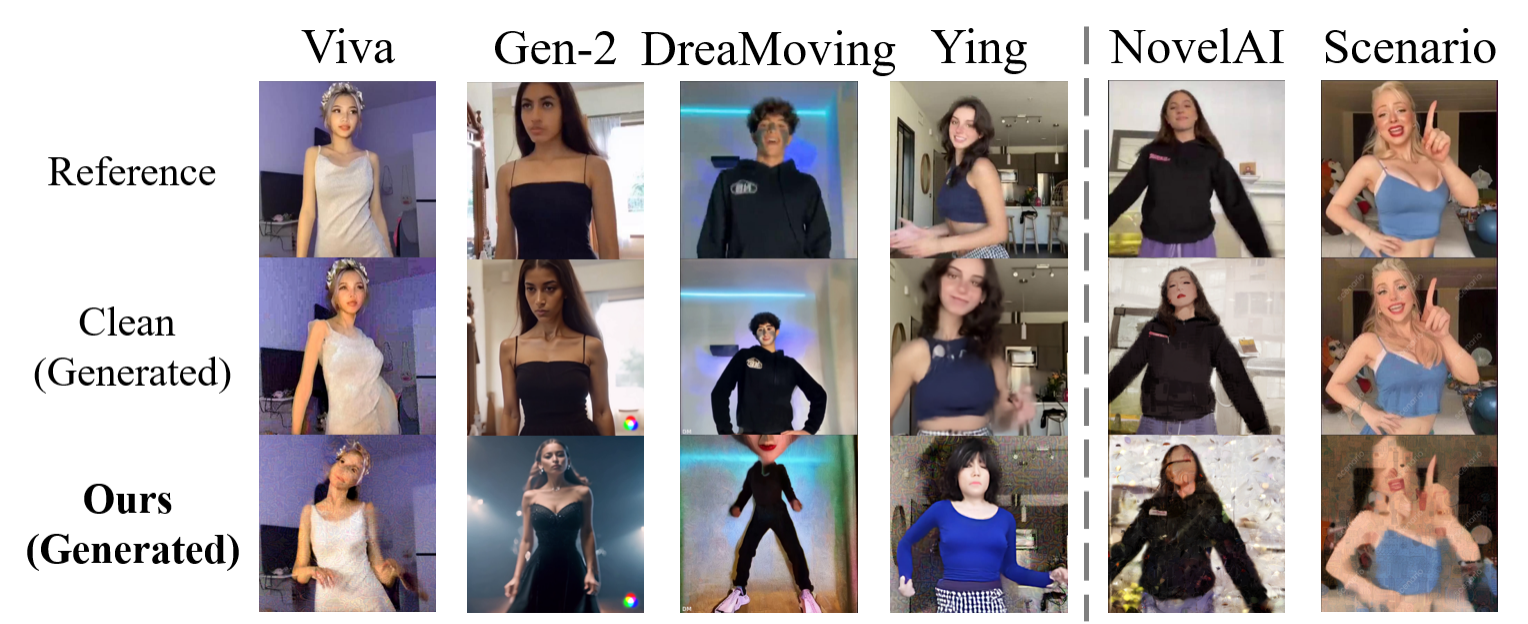}
    \caption{Qualitative results on commercial services.}
    \label{fig:qualitative results on commercial services}
\end{figure}

\begin{table}[t]
\caption{Quantitative results on various image-to-video and image-to-image methods.}
\label{tab:quantitative results on various image-to-video and image-to-image methods}
\centering
\scriptsize
\tabcolsep=0.08cm
\renewcommand\arraystretch{0.974}
\begin{tabular}{cccccccc}
    \toprule
    \textbf{Task} &\multicolumn{2}{c}{\textbf{Method}} &\textbf{CLIP-I$\downarrow$} &\textbf{DINO$\downarrow$}  &\textbf{SSIM$\downarrow$} &\textbf{LPIPS$\uparrow$} &\textbf{FID$\uparrow$}\\
    \midrule
    \multirow{8}{*}{\rotatebox{90}{\textbf{Image-to-Video}}} &\multirow{2}{*}{\textbf{PIA~\cite{zhang2024pia}}} & \textbf{Clean} & 0.752 & 0.768 & 0.573 & 0.519 & 142.69 \\
    & &\textbf{Protect} & 0.655 & 0.464 & 0.409 & 0.693 & 275.74\\
    \cmidrule{2-8}
    &\multirow{2}{*}{\textbf{I2VGen-XL~\cite{zhang2023i2vgen}}} & \textbf{Clean} & 0.779 & 0.730 & 0.467 & 0.560 & 158.72 \\
    & &\textbf{Protect} & 0.641 & 0.414 & 0.284 & 0.690 & 300.19 \\
    \cmidrule{2-8}
    &\multirow{2}{*}{\textbf{SVD~\cite{blattmann2023stable}}} & \textbf{Clean} & 0.799 & 0.805 & 0.540 & 0.476 & 124.69 \\
    & &\textbf{Protect} & 0.658 & 0.377 & 0.317 & 0.683 & 321.32 \\
    \cmidrule{2-8}
    &\multirow{2}{*}{\textbf{AnimateDiff~\cite{guo2024animatediff}}} & \textbf{Clean} & 0.752 & 0.831 & 0.655 & 0.447 & 120.18 \\
    & &\textbf{Protect} & 0.596 & 0.416 & 0.359 & 0.642 & 279.09\\
    \midrule
    \multirow{8}{*}{\rotatebox{90}{\textbf{Image-to-Image}}} &\multirow{2}{*}{\textbf{SDEdit~\cite{meng2022sdedit}}} & \textbf{Clean} & 0.733 & 0.841 & 0.609 & 0.442 & 140.98 \\
    & &\textbf{Protect} & 0.599 & 0.535 & 0.305 & 0.634 & 264.35 \\
    \cmidrule{2-8}
    &\multirow{2}{*}{\textbf{LEDITS++~\cite{brack2024ledits++}}} & \textbf{Clean} & 0.823 & 0.853 & 0.735 & 0.288 & 119.50 \\
    & &\textbf{Protect} & 0.682 & 0.591 & 0.488 & 0.626 & 222.32 \\
    \cmidrule{2-8}
    &\multirow{2}{*}{\textbf{DiffEdit~\cite{couairon2023diffedit}}} & \textbf{Clean} & 0.788 & 0.843 & 0.738 & 0.267 & 180.22 \\
    & &\textbf{Protect} & 0.665 & 0.430 & 0.428 & 0.649 & 375.78 \\
    \cmidrule{2-8}
    &\multirow{2}{*}{\textbf{InstructPix2Pix~\cite{brooks2023instructpix2pix}}} & \textbf{Clean} & 0.773 & 0.772 & 0.653 & 0.391 & 156.13 \\
    & &\textbf{Protect} & 0.675 & 0.529 & 0.420 & 0.686 & 275.54 \\
    \bottomrule	
\end{tabular}
\end{table}

As shown in Table~\ref{tab:quantitative results on various image-to-video and image-to-image methods}, \tool~demonstrates excellent protection performance, resulting in significant degradation in the quality of generated videos and images. For the four image-to-video methods, the average CLIP-I, DINO, and SSIM decrease from 0.771, 0.784, and 0.559 to 0.638, 0.418, and 0.342, and the average LPIPS and FID increase from 0.501 and 136.57 to 0.677 and 294.09, respectively. While for the four image-to-image methods, the average CLIP-I, DINO, and SSIM decrease from 0.779, 0.827, and 0.684 to 0.655, 0.521, and 0.410, and the average LPIPS and FID increase from 0.347 and 149.21 to 0.649 and 284.50, respectively. Visualization results are shown in~Figure~\ref{fig:qualitative results on image-to-video methods} and Figure~\ref{fig:qualitative results on image-to-image methods}, where both the generated video frames and images exhibit poor quality. These results further highlight the effectiveness and transferability of \tool. The features contained in the reference image are effectively and comprehensively disrupted by the protective perturbation from the perspective of various LDMs used in these generation methods, which utilize different architectures and are trained on different datasets.

\begin{table}[ht]
\caption{Quantitative results on various commercial services.}
\label{tab:quantitative results on various commercial services}
\centering
\scriptsize
\tabcolsep=0.08cm
\renewcommand\arraystretch{1}
\begin{tabular}{cccccccc}
    \toprule
    \textbf{Task} &\multicolumn{2}{c}{\textbf{Commercial Service}} &\textbf{CLIP-I$\downarrow$} &\textbf{DINO$\downarrow$}  &\textbf{SSIM$\downarrow$} &\textbf{LPIPS$\uparrow$} &\textbf{FID$\uparrow$}\\
    \midrule
    \multirow{8}{*}{\rotatebox{90}{\textbf{Image-to-Video}}} &\multirow{2}{*}{\textbf{Viva~\cite{hidream2023viva}}} & \textbf{Clean} & 0.880 & 0.888 & 0.717 & 0.359 & 82.94 \\
    & &\textbf{Protect} & 0.652 & 0.486 & 0.418 & 0.642 & 308.24 \\
    \cmidrule{2-8}
    &\multirow{2}{*}{\textbf{Gen-2~\cite{runway2023gen}}} & \textbf{Clean} & 0.778 & 0.752 & 0.633 & 0.477 & 150.84 \\
    & &\textbf{Protect} & 0.661 & 0.604 & 0.543 & 0.655 & 236.08 \\
    \cmidrule{2-8}
    &\multirow{2}{*}{\textbf{DreaMoving~\cite{alibaba2023dreamoving}}} & \textbf{Clean} & 0.805 & 0.764 & 0.539 & 0.551 & 138.50 \\
    & &\textbf{Protect} & 0.664 & 0.563 & 0.470 & 0.661 & 215.73 \\
     \cmidrule{2-8}
    &\multirow{2}{*}{\textbf{Ying~\cite{zhipu2024ying}}} & \textbf{Clean} & 0.847 & 0.729 & 0.535 & 0.605 & 170.39 \\
    & &\textbf{Protect} &  0.677 & 0.472 & 0.407 & 0.681 & 278.24 \\
    \midrule
    \multirow{4}{*}{\rotatebox{90}{\makecell{\textbf{Image-to-} \\ \textbf{Image}}}}
    &\multirow{2}{*}{\textbf{NovelAI~\cite{anlatan2023novelai}}} & \textbf{Clean} & 0.710 & 0.724 & 0.654 & 0.493 & 186.74 \\
    & &\textbf{Protect} & 0.652 & 0.532 & 0.518 & 0.591 & 243.71 \\
    \cmidrule{2-8}
    &\multirow{2}{*}{\textbf{Scenario~\cite{scenario2023scenario}}} & \textbf{Clean} & 0.803 & 0.870 & 0.702 & 0.360 & 141.91 \\
    & &\textbf{Protect} & 0.663 & 0.466 & 0.493 & 0.626 & 335.85 \\
    \bottomrule	
\end{tabular}
\end{table}

\noindent\textbf{Results on Commercial Services.} We further evaluate the performance of \tool~on six real-world commercial services. To effectively defend against these closed-source generative models in a black-box manner, we set the budget $\eta$ to $32/255$ to generate protective perturbation. The results are presented in Table~\ref{tab:quantitative results on various commercial services}. \tool~significantly degrades the quality of generated videos and images, with the average CLIP-I, DINO, and SSIM decreasing from 0.804, 0.788, and 0.630 to 0.662, 0.521, and 0.475, and the average LPIPS and FID increasing from 0.474 and 145.22 to 0.643 and 269.64, respectively. As shown in Figure~\ref{fig:qualitative results on commercial services}, generated videos and images suffer from poor quality, such as mismatched human appearance (Gen-2, DreaMoving, and Ying) and distorted visuals (Viva, NovelAI, and Scenario). These results validate that \tool~effectively safeguards portrait and privacy rights, showcasing its capability for real-world applications in preventing human image misuse.

\subsection{Ablation Study}
\label{subsec:ablation study}

\begin{figure}[htbp]
   \centering
    \includegraphics[width=0.393\textwidth]{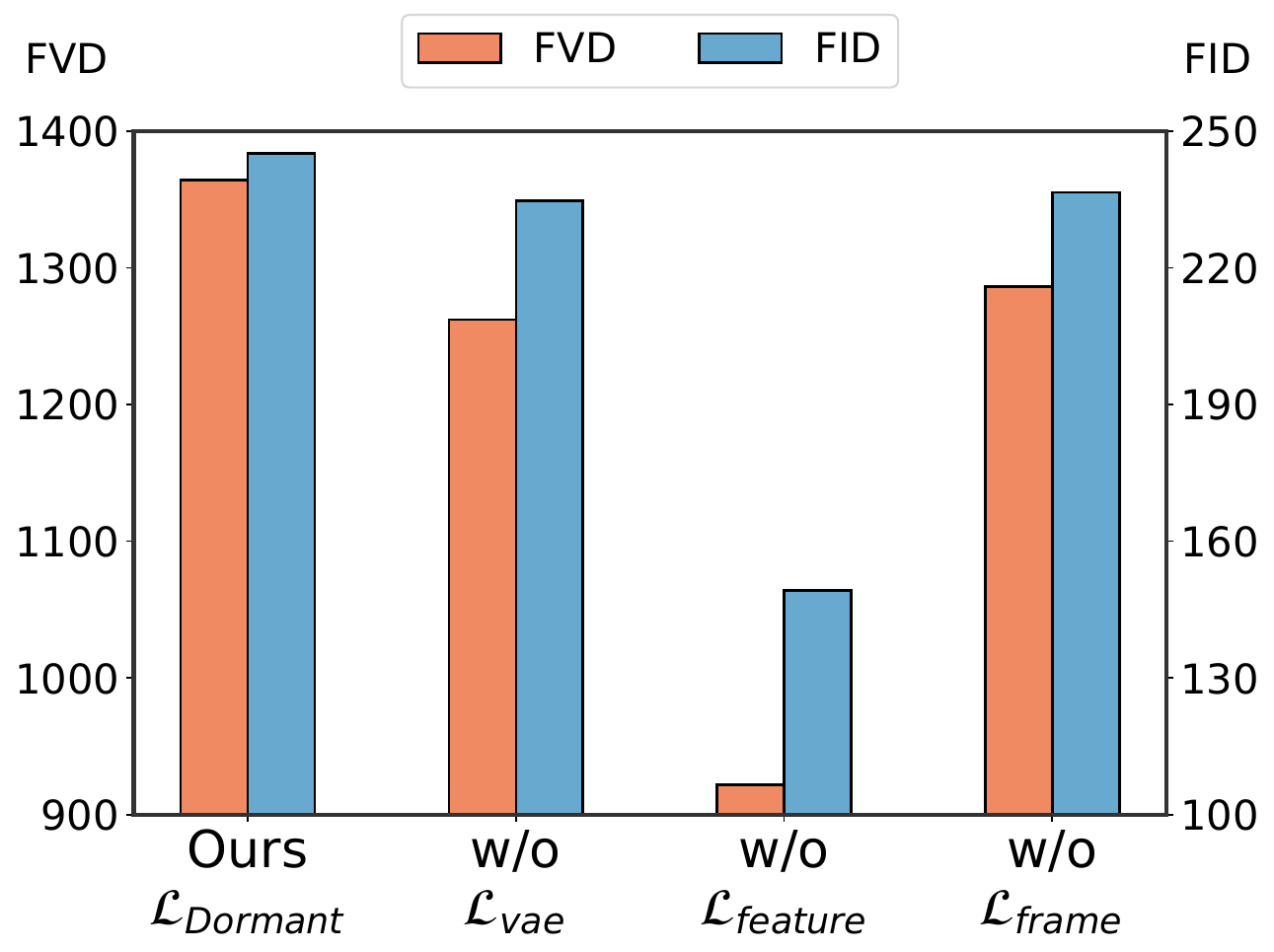}
    \caption{Ablation study on the proposed objectives.}
    \label{fig:ablation study on the proposed objectives}
\end{figure}

\noindent\textbf{Impact of Proposed Objectives.} To study the impact of $\mathcal{L}_{vae}$ and $\mathcal{L}_{feature}$ (feature misextraction), $\mathcal{L}_{frame}$ (frame incoherence), we conduct experiments by removing each corresponding loss while keeping the rest of $\mathcal{L}_\tool$ unchanged. As shown in Figure~\ref{fig:ablation study on the proposed objectives}, the absence of any of these three losses results in a degradation of the protection effect, highlighting their essential roles in the overall effectiveness of the protection. Notably, $\mathcal{L}_{feature}$ has the most significant impact, which is designed to induce misextraction of appearance features from the reference image using CLIP and ReferenceNet. This result aligns with our analysis in Section~\ref{subsec:protection methods against LDM}, where we emphasize that the main usage of the reference image in pose-driven human image animation is to extract its appearance features to guide video generation. Consequently, as shown in Table~\ref{tab:quantitative comparisons with baseline protections}, existing protections, which lack specific designs to effectively disrupt these features, exhibit limited efficacy.

We further study the impact of using other feature extractors in $\mathcal{L}_{feature}$. As mentioned in Section~\ref{subsec:feature misextraction}, we use the CLIP image encoder to extract semantic features and ReferenceNets to capture fine-grained details from the image. We replace our design of CLIP + ReferenceNets with DINO~\cite{caron2021emerging} + BLIP image encoder~\cite{li2022blip}, yielding FVD and FID values of 1080.09 and 153.55 for the generated videos. This result outperforms most baseline protections (except Glaze) but falls short of \tool~by a noticeable margin, further highlighting the importance of inducing misextraction of appearance features and also the superiority of our design for feature extractors.

\begin{figure}[hb]
  \centering
  \begin{subfigure}[b]{0.495\linewidth}
        \centering
        \includegraphics[width=\textwidth]{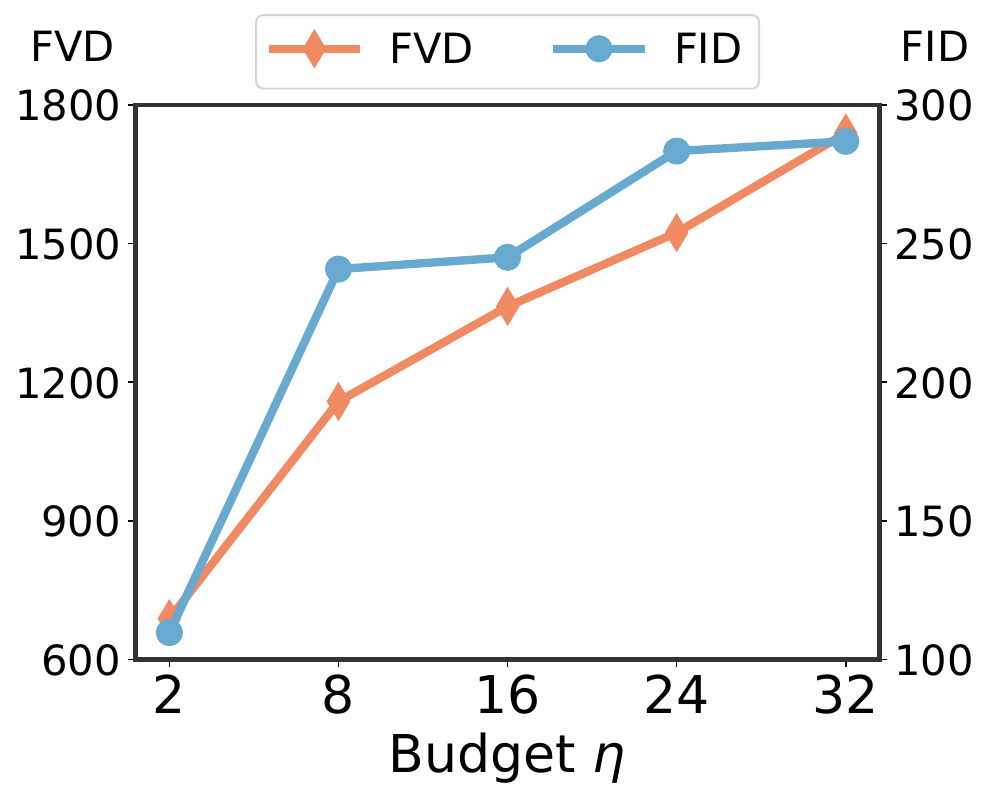}
        \caption{Perturbation budget $\eta$}
        \label{fig:impact of budget}
  \end{subfigure}
  \hfill
  \begin{subfigure}[b]{0.495\linewidth}
        \centering
        \includegraphics[width=\textwidth]{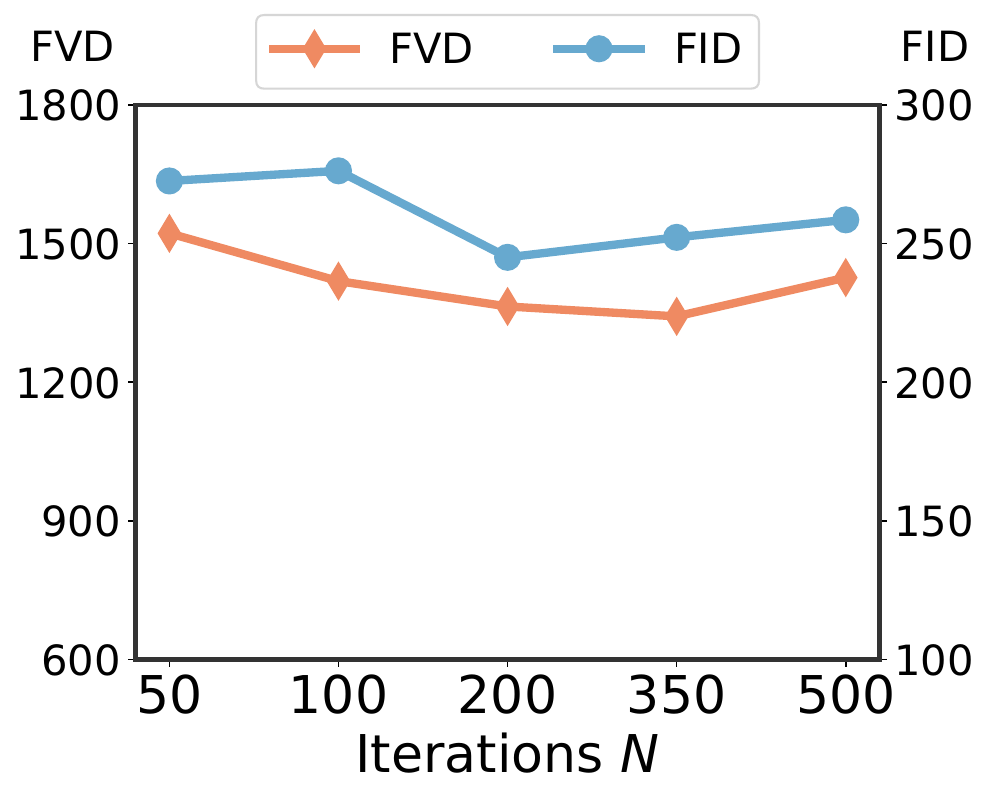}
        \caption{PGD iterations $N$}
        \label{fig:impact of iterations}
  \end{subfigure}
  \caption{Impact of budget $\eta$ and iterations $N$.}
  \label{fig:impact of budget and iterations}
\end{figure}

\noindent\textbf{Impact of Perturbation Budget $\eta$ and PGD Iterations $N$.} We conduct experiments by varying $\eta$ from $2/255$ to $32/255$ and $N$ from $50$ to $500$, to investigate their impact on protection performance. The results are presented in Figure~\ref{fig:impact of budget and iterations}. As the perturbation budget $\eta$ increases, the protection effect of \tool~becomes progressively stronger, and $\eta=8/255$ can already significantly degrade the quality of generated videos, resulting in FVD and FID values of 1158.46 and 240.83, respectively. This provides users with the flexibility to balance invisibility and effectiveness according to their needs. The increase in PGD iterations $N$ initially causes the protection effect to decrease and then increase, possibly due to overfitting, which leads the optimization to fall into poor local maxima and then escape. Optimizing for 50 iterations is sufficient to significantly degrade the quality of generated videos, with FVD and FID being 1522.35 and 272.61, thereby further reducing the time cost required for protection. Impact of pose repeats $F$ and decay factor $\mu$ is provided in Appendix~\ref{sec:more ablation study}. 

\section{Discussion}
\label{sec:discussion}

\noindent\textbf{Adaptive Attack.} We propose an adaptive attack under the assumption that the attacker has access to multiple protected images of the victim. While animation methods typically take a single reference image as input to generate videos, the attacker can utilize these images to craft a refined reference. Inspired by~\cite{passananti2024disrupting} which reveals Glaze’s vulnerability to linear interpolation and pixel averaging, we design a strategy to create a purified image from multiple protected images. Specifically, we apply linear interpolation to adjacent pairs of five protected images spaced five frames apart, followed by pixel averaging of the four interpolated results to produce the final image for video generation, with the resulting FID-FVD, FVD, LPIPS, FID, PSNR, and SSIM on the TikTok dataset being 69.18, 728.89, 0.474, 162.50, 17.06, and 0.641, respectively. While the protective effect is partially reduced, the generated videos remain of noticeably low quality. More adaptive attacks are provided in Appendix~\ref{sec:more adaptive attacks}.

\noindent\textbf{Application Scenarios.} \tool~applies protective perturbations to human images to prevent potential unauthorized usage in pose-driven animation methods. It is primarily targeted at users who: 1) may place human images in uncontrollable environments like social media or the web; and 2) prioritize privacy and portrait rights and accept sacrificing a bit of image quality for protection. \tool~may not be directly suitable for legitimate use cases that involve processing or editing original images for valid purposes, such as photo retouching or authorized content creation.

\begin{figure}[htbp]
  \centering
  \begin{subfigure}[b]{0.495\linewidth}
        \centering
        \includegraphics[width=\textwidth]{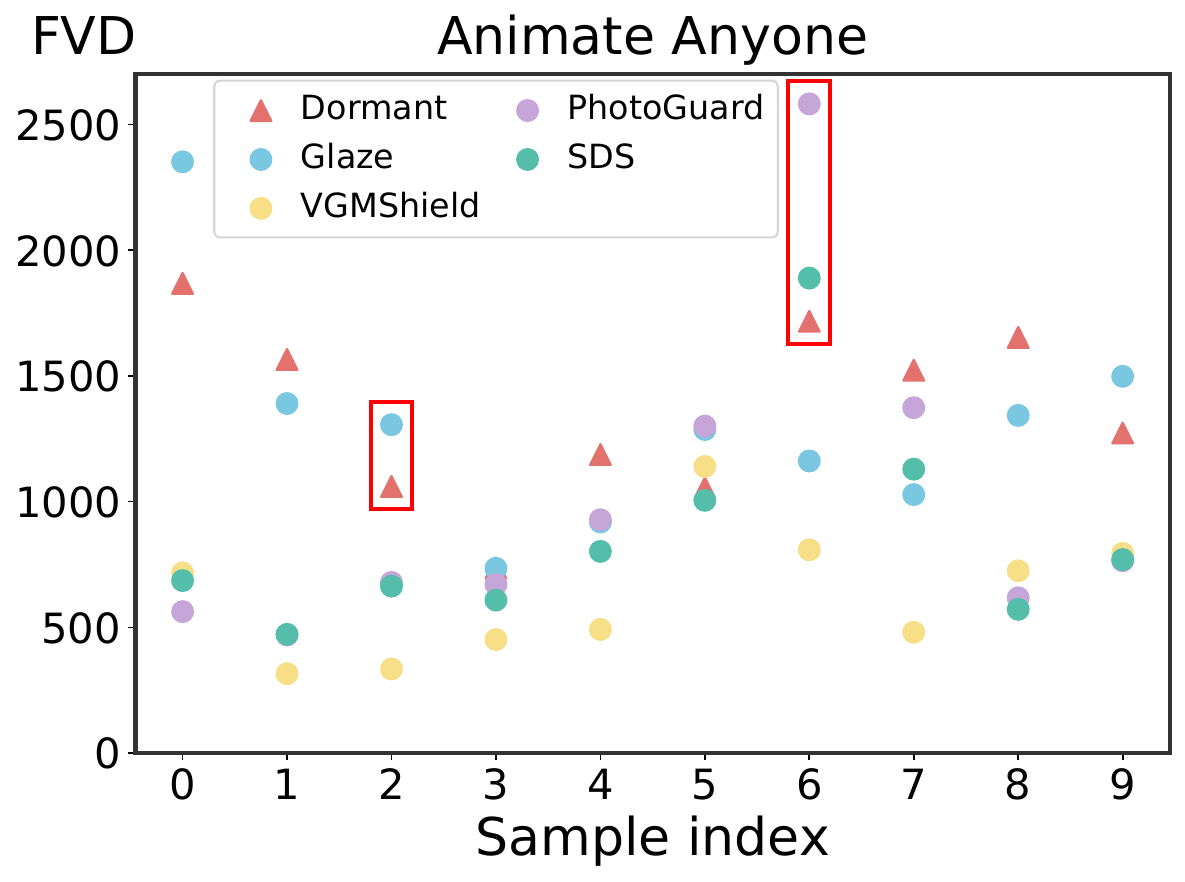}
        \caption{Results of Animate Anyone}
        \label{fig:results of Animate Anyone}
  \end{subfigure}
  \hfill
  \begin{subfigure}[b]{0.495\linewidth}
        \centering
        \includegraphics[width=\textwidth]{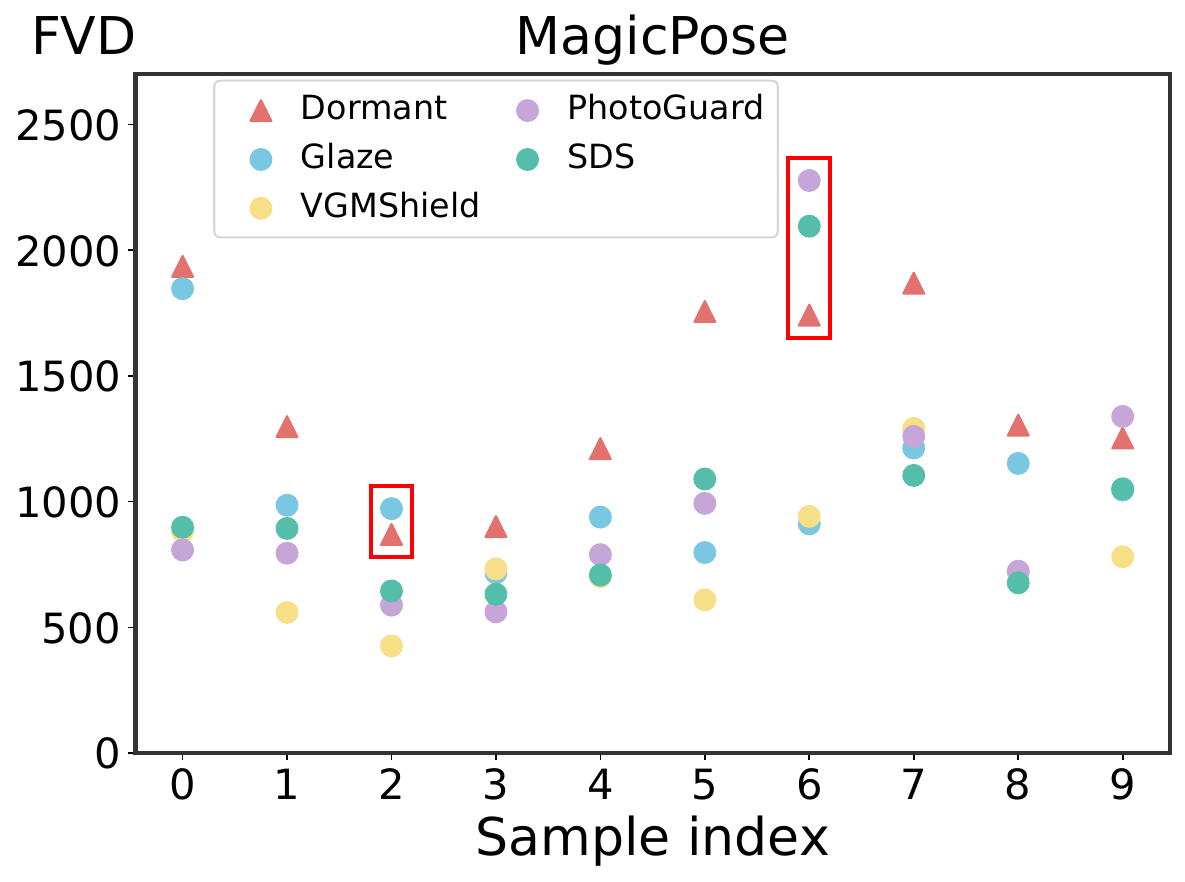}
        \caption{Results of MagicPose}
        \label{fig:results of MagicPose}
  \end{subfigure}
  \caption{Results of experiments on Animate Anyone and MagicPose, with red boxes marking samples where baseline protections perform better~across both animation methods.}
  \label{fig:results of Animate Anyone and MagicPose}
\end{figure}

\noindent\textbf{Anomalous Cases.} We examine cases where baseline protections outperform \tool~across at least four animations on the TikTok dataset. While \tool~exhibits sufficient effectiveness, these methods yield even better protective effects on certain samples. Specifically, Glaze performs better on sample 002, producing highly exaggerated colors and visible speckles in the generated videos. PhotoGuard and SDS perform better on sample 006 which features sharply contrasting colors, by lightening the colors and blurring sharpness and details in the generated videos. We present the FVD for experiments on Animate Anyone and MagicPose as examples in Figure~\ref{fig:results of Animate Anyone and MagicPose}, and qualitative results in Figure~\ref{fig:visualization not outperform} in Appendix~\ref{sec:visualized results of anomalies}.

We also examine failure cases where \tool~performs relatively poorly against the commercial service Gen-2 and the latest Gen-3. As shown in Figure~\ref{fig:visualization perform relatively poorly} in Appendix~\ref{sec:visualized results of anomalies}, \tool~mainly induces identity mismatch but introduces minimal distortion, and the generated videos remain realistic. Image-to-video models applied in commercial services are typically trained on larger, higher-quality datasets with more parameters. Exploring how to get feedback from these advanced models via black-box access to improve \tool's transferability would be meaningful for future work.

\noindent\textbf{Limitations and Future Works.} A key limitation of \tool~is that its effectiveness is validated based on empirical evaluation, rather than being provably secure. Additionally, \tool~faces the challenge of being future-proof, as image-to-video generation is continuously evolving. While \tool~has shown effectiveness and is expected to remain effective against existing feature extractors used in current open-source SOTA animation methods that are driven by pose, its performance against future paradigms in feature extraction and different video generation techniques guided by other modal conditions requires ongoing validation and improvements. Future work could explore the use of expert models to specifically extract disentangled image features (\eg, human face, body posture, and background) to induce feature misextraction, and integrate guidance from additional modalities to create frame incoherence during optimization. 

\section{Related Work}
\label{sec:related work} 

\noindent\textbf{LDM for Unauthorized Image Usage.} Latent diffusion models~\cite{rombach2022high} have demonstrated a remarkable capability to create authentic-looking images, which also raises significant concerns about unauthorized image usage. Malicious attackers can easily mimic an artist's style~\cite{gal2023an,hu2022lora,kumari2023multi,ruiz2023dreambooth} or edit images into new contexts~\cite{brack2024ledits++,brooks2023instructpix2pix,couairon2023diffedit,meng2022sdedit,zhang2023adding} without consent. Beyond image generation, unauthorized images can also be used to create fabricated videos~\cite{blattmann2023stable,guo2024animatediff,zhang2023i2vgen,zhang2024pia}. In particular, pose-driven human image animation~\cite{chang2024magicpose,fang2023dance,hu2024animate,huang2024magicfight,karras2023dreampose,lei2024comprehensive,peng2024controlnext,tong2024musepose,wang2024disco,wang2024unianimate,xia2024musev,xu2024magicanimate,zhu2024champ,zhang2024mimicmotion,wang2024vividpose} can generate malicious videos by animating human images using pose sequences as guidance.

\noindent\textbf{Protection Methods and Countermeasures.} Previous protection methods apply protective perturbations to images to prevent unauthorized usage in DNN-based recognition~\cite{li2019hiding,shan2020fawkes,cherepanova2021lowkey} and GAN-based manipulation~\cite{li2023unganable,yeh2020disrupting,huang2021initiative}. However, these methods have shown limited effectiveness against advanced LDMs due to their distinct backbone networks~\cite{liang2023adversarial,shan2023glaze}. To defend against LDM-based text/image-to-image generation, recent research has proposed protections targeting customization techniques~\cite{zheng2023improving,van2023anti,liu2024metacloak}, VAE~\cite{shan2023glaze,salman2023raising,ye2024duaw}, and denoising UNet~\cite{liang2023adversarial,xue2024toward,zhang2023robustness}. The only concurrent work addressing unauthorized LDM-based image-to-video generation is VGMShield~\cite{pang2024vgmshield}, which targets the image and video encoders of SVD. There also exists countermeasures~\cite{an2024rethinking,cao2024impress,honig2025adversarial,liu2024investigating,zhao2024can} that scrutinize the effectiveness and robustness of current protection methods, proposing various techniques to circumvent them by purifying the added protective perturbation.
\section{Conclusion}
\label{sec:conclusion}

In this paper, we propose \tool, a novel approach designed to prevent unauthorized image usage in pose-driven human image animation. By applying the protective perturbation to the human image, \tool~effectively degrades the quality of generated videos, resulting in mismatched appearances, visual distortions, and frame inconsistencies. To optimize the protective perturbation, we develop specific objective functions aimed at inducing misextraction of appearance features and causing incoherence among generated video frames. Experimental results demonstrate the effectiveness, transferability, and robustness of \tool, serving as a powerful tool for safeguarding portrait and privacy rights.

\section*{Acknowledgments}
\label{sec:acknowledgments}

We thank our shepherd and all the anonymous reviewers for their constructive feedback. The IIE authors are supported in part by NSFC (92270204, U24A20236), CAS Project for Young Scientists in Basic Research (Grant No. YSBR-118). 

\section*{Ethics Considerations}
\label{sec:ethics considerations}

\tool~is a defense approach designed to prevent image misuse in pose-driven human image animation methods, thereby safeguarding individuals' rights to portrait and privacy. To ensure fair and reproducible evaluations of the performance of \tool, we conducted experiments on commonly used public datasets, following the setups outlined in prior animation methods~\cite{chang2024magicpose,wang2024disco,zhu2024champ,xu2024magicanimate}. All experimental data, including the fake videos generated during the experiments, were stored on private, secure servers and are strictly intended for academic research purposes only; they will not be shared beyond their intended scope. 

During the research process, three experiments posed potential risks of unauthorized exposure of generated fake videos: the human study, the GPT-4o study, and experiments on commercial services. These experiments used the TikTok test set~\cite{jafarian2021learning,wang2024disco}, which permits research purposes. Although this publicly available dataset is widely used in many studies, obtaining explicit consent from the individuals depicted in the photos remains challenging. An alternative approach could have involved collecting and using data from volunteers who could explicitly consent. However, the terms of service for video generation platforms typically allow the use of user inputs and outputs for training and improving their AI models. While publicly available data may have already been integrated into these platforms’ model training, uploading newly collected images from consenting individuals would introduce their data for the first time, potentially exposing them to additional privacy risks due to uncontrolled usage. To mitigate these risks and ensure fairness and reproducibility in our evaluation, we chose to use publicly available data.

Meanwhile, we remained vigilant about the potential exposure of generated fake videos and implemented precautionary measures to minimize risks. All participants in the human study were required to consent to not disseminate any content from the questionnaire, which was accessible for a limited time. Before uploading data to GPT-4o and video generation websites, we configured our account privacy settings to the strictest levels, including disabling ``Chat History \& Training'' in GPT-4o, setting repositories to private, and upgrading to memberships with enhanced privacy features, \etc~During the experiments, we documented the setups for reproducibility, downloaded and stored the generated results on private servers, and cleared all data from the accounts upon completion. All activities were conducted in full compliance with the usage rules and regulations of the respective platforms.

We conducted a human study to evaluate the performance of \tool~from the perspective of human perception. The study was reviewed and received IRB approval at our institute. Before providing consent, participants were fully informed of the study’s purpose, procedures, potential risks and benefits, and lottery-based compensation. They were also warned that the questionnaire might include unfiltered videos that could be disturbing, and were shown sample videos upon request if they expressed concern. Participants were allowed to withdraw from the study at any point, with any data they provided excluded from the analysis. Additionally, participants were required to consent to not disseminate any content from the questionnaire. Explicit consent was obtained through a digital agreement before participants gained access to the questionnaire, which was available for a limited duration. All participants remained anonymous. Demographic data, including gender, age, education, expertise, and familiarity, were collected solely for research purposes through single-choice questions with generalized options, including a ``prefer not to say'' option. No personally identifiable information was collected, and all data were securely and privately stored.

\section*{Open Science}
\label{sec:open science}

We are committed to the principles of open science and have released our artifacts, including the source code of \tool~and the evaluated test data at\url{https://zenodo.org/records/14725876}. The latest version of the code will be maintained and updated at\url{https://github.com/Manu21JC/Dormant}. While we primarily present visualizations of experimental results via figures in this paper, we do not plan to provide public links to these generated fake videos to mitigate potential risks of exposure and misuse, as discussed in Ethics Considerations. These videos are available upon request for research purposes only, with the condition that requestors agree not to share the videos beyond their intended scope.

{\footnotesize 
\bibliographystyle{plain}
\bibliography{reference}}

\appendix
\begin{center}
   \LARGE \textbf{Appendix} 
\end{center}

\section{Details of Evaluated Methods}
\label{sec:details of evaluated methods}

In this section, we provide detailed introductions to the evaluated pose-driven human image animation methods, baseline protection methods, transformations, purifications, image-to-video methods, and image-to-image methods, which are mainly presented in chronological order.

\noindent\textbf{Pose-driven Human Image Animation Methods.} To leverage the prior knowledge of image diffusion models for human image animation, MagicPose~\cite{chang2024magicpose} introduces two plug-in sub-modules for SD, including Appearance Control Model to provide appearance guidance and Pose ControlNet to offer pose guidance. MagicAnimate~\cite{xu2024magicanimate} inflates SD's 2D UNet to 3D temporal UNet by inserting temporal attention layers to encode temporal information and proposes a novel appearance encoder (a trainable copy of SD's 2D UNet) to extract dense visual features from the reference image. Animate Anyone~\cite{hu2024animate} designs ReferenceNet to extract detailed features from the reference image and Pose Guider to encode motion control signals. Champ~\cite{zhu2024champ} incorporates rendered depth images, normal maps, and semantic maps obtained from SMPL sequences, along with skeleton-based motion to guide video generation. MuseV~\cite{xia2024musev} introduces a novel Visual Conditioned Parallel Denoising scheme to support infinite-length video generation. MusePose~\cite{tong2024musepose} proposes a pose align algorithm to align arbitrary dance videos with arbitrary reference images. UniAnimate~\cite{wang2024unianimate} leverages a unified video diffusion model to map the reference image, posture guidance, and noise video into a common feature space. ControlNeXt~\cite{peng2024controlnext} implements human image animation on SVD by introducing a lightweight convolutional module to extract control features and replacing zero convolution with cross normalization to achieve fast and stable training convergence.

\noindent\textbf{Baseline Protections.} Glaze~\cite{shan2023glaze} applies carefully computed perturbations (``style cloaks'') to artworks, shifting their representation in VAE's feature space towards a target art style, such that text-to-image models trained on cloaked images would learn the incorrect representation of the artist's style. PhotoGuard~\cite{salman2023raising} optimizes the perturbation by attacking either the VAE encoder or the entire diffusion process of the LDM. AntiDB~\cite{van2023anti} optimizes the perturbation by attacking the learning process of DreamBooth and introduces Fully-trained Surrogate Model Guidance and Alternating Surrogate and Perturbation Learning to solve this bi-level optimization. Mistv2~\cite{zheng2023improving} reveals the dynamics of adversarial attacks on LDM and improves performance by attacking with consistent score-function errors. SDS~\cite{xue2024toward} proposes Score Distillation Sampling to double the optimization speed and reduce memory occupation. VGMShield~\cite{pang2024vgmshield} prevents image misuse in image-to-video generation by deceiving both the image and video encoders of SVD into misinterpreting the image.

\noindent\textbf{Transformations.} Gaussian noise adds noise to the image, following a probability density function based on a Gaussian distribution, characterized by a mean and standard deviation. Gaussian blur convolves the image with a Gaussian function, replacing each pixel’s value with a weighted average of its neighboring pixels, where the weights are determined by a Gaussian distribution. Median blur applies a median filter to the image, replacing each pixel's value with the median value of its neighboring pixels within a defined kernel size. JPEG compression compresses an image to a binary JPEG file by: converting the RGB image to the YCbCr color space, subsampling the chrominance channels, dividing the image into $8 \times 8$ pixel blocks, applying the discrete cosine transform to each block, quantizing the frequency coefficients, and performing Huffman coding to build the JPEG file. Bit squeeze~\cite{xu2018feature} reduces the color bit depth of each pixel to mitigate adversarial examples while maintaining human recognition.

\noindent\textbf{Purifications.} DiffPure~\cite{nie2022diffusion} utilizes diffusion models for adversarial purification by first diffusing the adversarial example with a small amount of noise through a forward process, and then recovering the clean image via a reverse process. DDSPure~\cite{carlini2023certified} employs pre-trained diffusion models as the denoiser for Denoised Smoothing to improve certified adversarial robustness to $L_{2}$-norm bounded perturbations. Impress~\cite{cao2024impress} purifies the protected image by optimizing both the LPIPS-based similarity loss and the consistency loss during reconstruction using the VAE encoder and decoder. GrIDPure~\cite{zhao2024can} generates the final purified image by iterating the following steps multiple times: dividing the protected image into multiple grids, purifying each grid using SDEdit, merging the purified grids by averaging any overlapping parts, and blending the merged purified image with the original protected image. Noisy Upscaling~\cite{honig2025adversarial} purifies the protected image by first applying Gaussian noise and then upscaling the noisy image using the Stable Diffusion Upscaler. Diffshortcut~\cite{liu2024investigating} purifies the protected image by using a face-oriented restoration model CodeFormer to restore human faces, and the Stable Diffusion Upscaler to purify non-face regions.

\noindent\textbf{Image-to-Video Methods.} AnimateDiff~\cite{guo2024animatediff} integrates motion dynamics into pre-trained text-to-image models, enabling the generation of animations by training a plug-and-play motion module on large-scale video datasets. I2VGen-XL~\cite{zhang2023i2vgen} proposes a cascaded video synthesis model that consists of two stages: the base stage guarantees coherent semantics and preserves content from input images using two hierarchical encoders, and the refinement stage enhances video details with additional text guidance and improves resolution. SVD~\cite{blattmann2023stable} identifies three stages for training video LDMs, including text-to-image pre-training, video pre-training on a large dataset at low resolution, and video fine-tuning on a small subset of higher-quality videos at high resolution. PIA~\cite{zhang2024pia} introduces a trainable module into the input layer of the text-to-image model, which takes the condition frame and inter-frame affinity as inputs to improve appearance alignment. 

\noindent\textbf{Image-to-Image Methods.} Given an input image with user guidance input (\eg, a stroke painting), SDEdit~\cite{meng2022sdedit} first adds noise to the input and then denoises the resulting image using the Stochastic Differential Equation to increase its realism. DiffEdit~\cite{couairon2023diffedit} identifies regions requiring edits to match the text query by contrasting predictions of a diffusion model conditioned on different text prompts and performs DDIM encoding and decoding with mask guidance to edit the image. InstructPix2Pix~\cite{brooks2023instructpix2pix} trains a conditional diffusion model on a large dataset of image editing examples generated by GPT-3 and SD, enabling intuitive image editing that can follow human-written instructions. LEDITS++~\cite{brack2024ledits++} proposes an inversion method to identify the noise vector of the input image, and edits the image by manipulating the noise estimate based on textual descriptions, using an implicit masking approach to semantically limit changes to relevant regions.

\section{Protection Generality}
\label{sec:protection generality}

\noindent\textbf{Results on Various Resolutions of Images.} As mentioned in Section~\ref{subsec:experimental setup}, our experiments are mainly conducted on $512 \times 512$ images. Here we investigate the protection performance of \tool~on four other resolutions: $\{256 \times 256, 256 \times 512, 512 \times 768, 768 \times 768\}$. As shown in Table~\ref{tab:quantitative results on various resolutions of images}, \tool~exhibits excellent protection performance across all image resolutions, with the average FID-VID, FVD, LPIPS and FID increasing from 53.83, 447.50, 0.31 and 81.01 to 215.96, 1628.55, 0.64 and 251.09, and the average PSNR and SSIM decreasing from 16.85 and 0.701 to 11.85 and 0.370, respectively. These results highlight the applicability of \tool~for protecting human images of different resolutions.

\begin{table}[htbp]
\caption{Quantitative results on various resolutions of images.}
\label{tab:quantitative results on various resolutions of images}
\centering
\scriptsize
\tabcolsep=0.08cm
\renewcommand\arraystretch{1}
\begin{tabular}{cccccccc}
    \toprule
    \multicolumn{2}{c}{\textbf{Resolution}} &\textbf{FID-VID$\uparrow$} &\textbf{FVD$\uparrow$} &\textbf{PSNR$\downarrow$} &\textbf{SSIM$\downarrow$} &\textbf{LPIPS$\uparrow$} &\textbf{FID$\uparrow$}\\
    \midrule
    \multirow{2}{*}{\textbf{256 $\bm{\times}$ 256}} & \textbf{Clean} & 56.23 & 462.36 & 16.04 & 0.625 & 0.304 & 94.07 \\
    & \textbf{Protect} & 260.26 & 2165.94 & 13.19 & 0.306 & 0.596 & 285.03 \\
    \midrule
    \multirow{2}{*}{\textbf{256 $\bm{\times}$ 512}} & \textbf{Clean} & 69.54 & 480.56 & 15.74 & 0.637 & 0.349 & 83.85 \\
    & \textbf{Protect} & 289.11 & 1769.24 & 12.06 & 0.322 & 0.640 & 292.09 \\
    \midrule
    \multirow{2}{*}{\textbf{512 $\bm{\times}$ 512}} & \textbf{Clean} & 41.61 & 362.75 & 17.76 & 0.741 & 0.282 & 65.00 \\
    & \textbf{Protect} & 162.87 & 1364.00 & 11.82 & 0.374 & 0.639 & 245.06 \\
    \midrule
    \multirow{2}{*}{\textbf{512 $\bm{\times}$ 768}} & \textbf{Clean} & 58.86 & 505.83 & 16.84 & 0.722 & 0.328 & 85.67 \\
    & \textbf{Protect} & 204.48 & 1419.64 & 11.62 & 0.403 & 0.653 & 243.46\\
    \midrule
    \multirow{2}{*}{\textbf{768 $\bm{\times}$ 768}} & \textbf{Clean} & 42.93 & 425.98 & 17.89 & 0.779 & 0.284 & 76.45 \\
    & \textbf{Protect} & 163.07 & 1423.93 & 10.58 & 0.441 & 0.650 & 189.82 \\
    \bottomrule	
\end{tabular}
\end{table}

\noindent\textbf{Results on Random References and Jump-cut Poses.} As mentioned in Section~\ref{subsec:experimental setup}, we use the first frame of each video as the reference image and extract pose sequences from the remaining frames for video generation. We further conduct two additional experiments to examine whether altering these settings would impact \tool's effectiveness. First, we randomly select a non-first frame from each video to serve as the reference image, with the resulting FID-FVD, FVD, LPIPS, FID, PSNR, and SSIM on the TikTok dataset being 162.83, 1352.56, 0.633, 225.50, 11.97, and 0.401, respectively. Second, we simulate jump-cut pose sequences by combining the first half of the pose sequence from one video with the second half from another, yielding CLIP-I, DINO, SSIM, LPIPS, and FID values of 0.609, 0.437, 0.404, 0.682, and 247.73, respectively. These results show that \tool's protection performance remains robust under both variations in the reference image and pose sequence settings.

\section{More Adaptive Attacks}
\label{sec:more adaptive attacks}

\noindent\textbf{Robustness against Robust Fine-tuning.} Inspired by~\cite{radiya-dixit2022data} which reveals that robust training with protected images and their corresponding ground-truth labels could defeat protections against face recognition models~\cite{shan2020fawkes,cherepanova2021lowkey}, we investigate whether this strategy would affect \tool's performance. In the image-to-video generation setting, it is hard for the attacker to robustly train a model from scratch due to the substantial computational resources and high-quality data required. Therefore, we focus on studying the impact of model fine-tuning on images protected by \tool. Specifically, we fine-tune the models of Animate Anyone using 1,000 protected images (with corresponding original images as targets) and 1,000 clean images from the TikTok dataset, yielding FID-FVD, FVD, LPIPS, FID, PSNR, and SSIM values of 93.52, 810.11, 0.432, 109.23, 15.01, and 0.615, respectively. While robust fine-tuning reduces \tool's effectiveness to some extent, the protective effect remains sufficient to ensure that the generated videos maintain low quality.

\noindent\textbf{Robustness against Targeted Removal.} We assume that the attacker has fully white-box access to \tool's mechanism, including the same pre-trained models and exact objective functions. Inspired by~\cite{an2024rethinking} which optimizes the purified image by pushing it in the opposite direction of the protected image and pulling it in the same direction of a visual reference, we implement this strategy in our proposed $\mathcal{L}_{\tool}$ and use purified images generated by DiffPure as visual references during optimization. The resulting FID-FVD, FVD, LPIPS, FID, PSNR, and SSIM values are 52.41, 485.01, 0.378, 97.73, 17.66, and 0.706, respectively. When compared to DiffPure's results presented in Table~\ref{tab:quantitative results against various purifications}, we find that white-box access does not lead to a noteworthy increase in the purification effect. While DiffPure remains the most effective among the evaluated purifications, future work could explore ways to further improve \tool's robustness against such diffusion-based adversarial purification methods.

\section{More Ablation Study}
\label{sec:more ablation study}

\begin{figure}[htbp]
  \centering
  \begin{subfigure}[b]{0.495\linewidth}
        \centering
        \includegraphics[width=\textwidth]{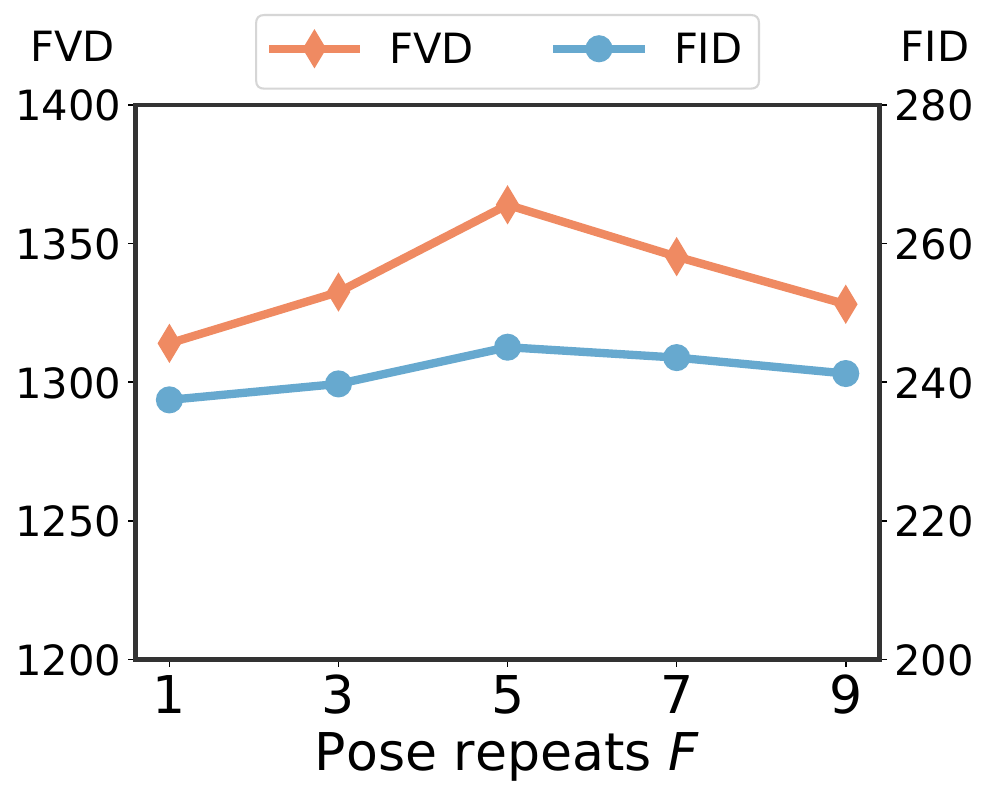}
        \caption{Pose repeats $F$}
        \label{fig:impact of pose repeats}
  \end{subfigure}
  \hfill
  \begin{subfigure}[b]{0.495\linewidth}
        \centering
        \includegraphics[width=\textwidth]{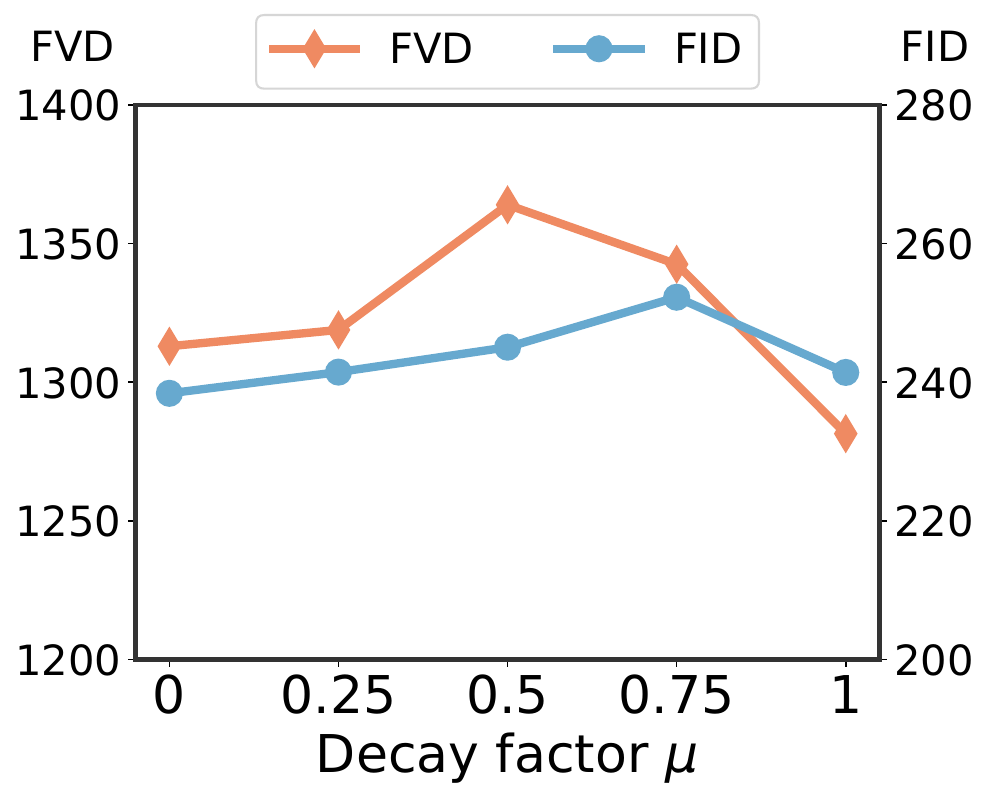}
        \caption{Decay factor $\mu$}
        \label{fig:impact of decay factor}
  \end{subfigure}
  \caption{Impact of pose repeats $F$ and decay factor $\mu$.}
  \label{fig:impact of pose repeats and decay factor}
\end{figure}

\noindent\textbf{Impact of Pose Repeats $F$ and Decay Factor $\mu$.} $F$ defines the length of the pose sequence used in $\mathcal{L}_{frame}$; $\mu$ controls the influence of previous gradients during each PGD step. As shown in Figure~\ref{fig:impact of pose repeats and decay factor}, both the increase in $F$ and $\mu$ cause the protection effect to first increase and then decline. While we optimize the protective perturbation to induce frame incoherence by maximizing the distance between each generated frame, increasing $F$ leads to more pairs of distances to compute, making the optimization more difficult and also increasing GPU memory usage. For the decay factor $\mu$, moderate memorization of previous gradients helps stabilize update directions and avoid poor local maxima, but excessive impact from previous gradients with a large $\mu$ would hinder the search for new update directions. We set the default values of $F$ and $\mu$ to 5 and 0.5, both within the sweet spot region.

\begin{figure}[htbp]
   \centering
   \includegraphics[width=0.47\textwidth]{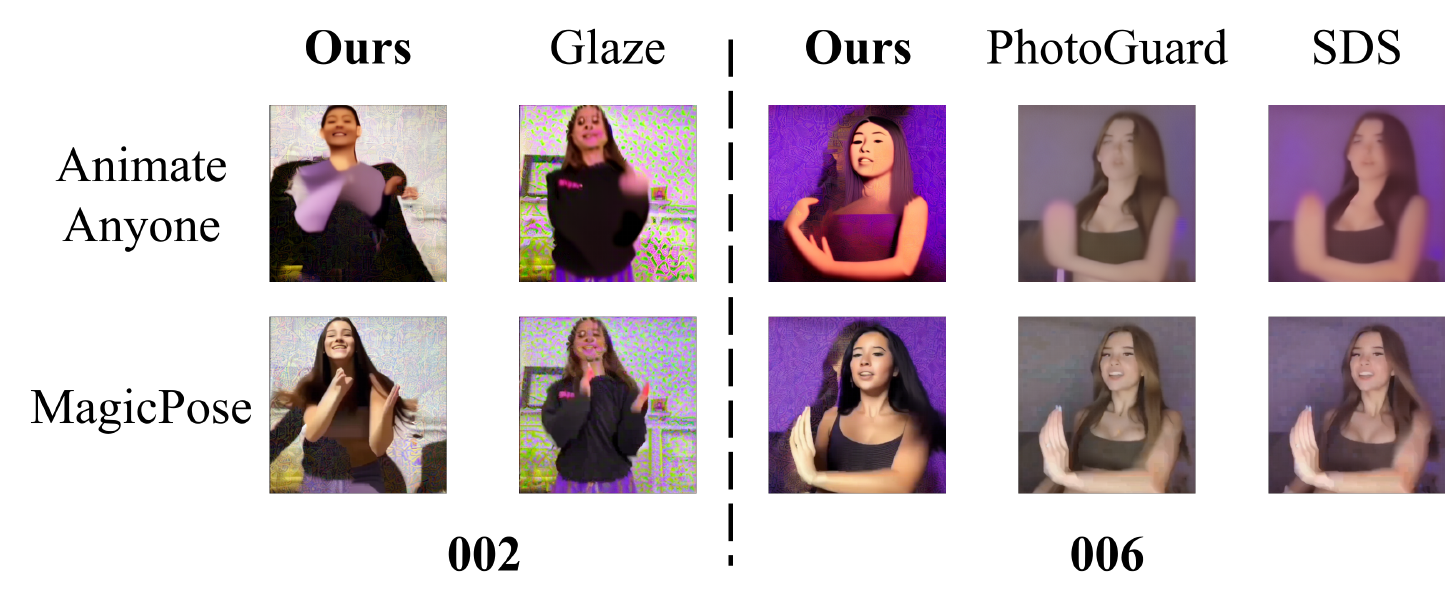}
    \caption{Visualizations of cases where \tool~does not outperform baseline protections.}
    \label{fig:visualization not outperform}
\end{figure}

\begin{figure}[htbp]
   \centering
   \includegraphics[width=0.47\textwidth]{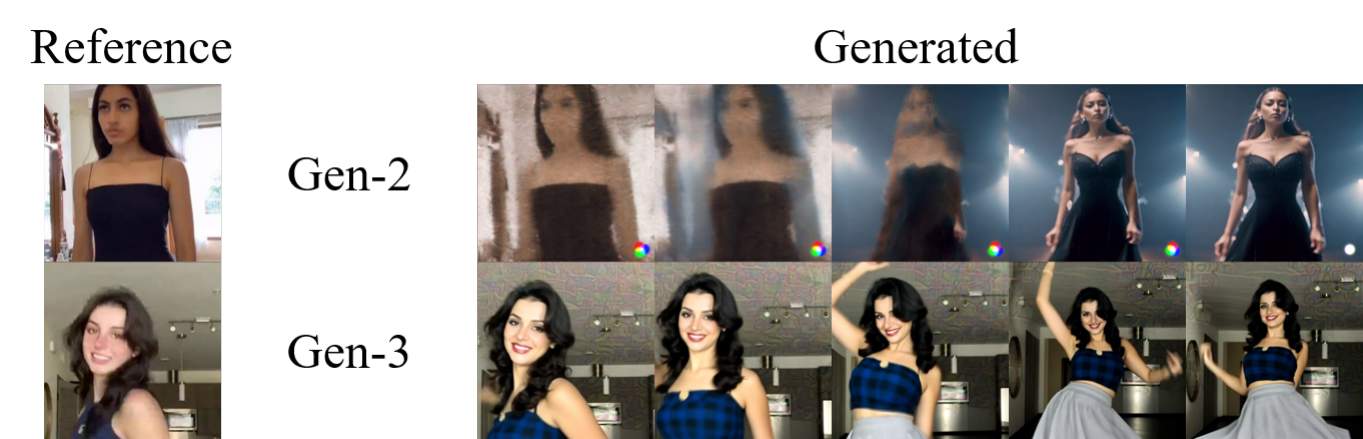}
    \caption{Visualizations of cases where \tool~performs relatively poorly.}
    \label{fig:visualization perform relatively poorly}
\end{figure}

\section{Visualized Results of Anomalies}
\label{sec:visualized results of anomalies}

As discussed in Section~\ref{sec:discussion}, we find that Glaze outperforms \tool~on sample 002, and PhotoGuard and SDS perform better on sample 006 of the TikTok dataset. Here we present the visualizations of generated videos in Figure~\ref{fig:visualization not outperform}. While \tool~has demonstrated sufficient effectiveness in these cases, these baseline protections perform even better on certain samples. Figure~\ref{fig:visualization perform relatively poorly} visualizes generated video frames against commercial services Gen-2 and Gen-3, where \tool~performs relatively poorly, mainly inducing identity mismatches without significant distortion.

\section{Details of Human and GPT-4o studies}
\label{sec:details of human and gpt-4o studies}

\subsection{Questionnaire in Human Study}
\label{subsec:questionnaire in human study}

We detail our questionnaire used in the human study, which is divided into two parts: demographics and video selection, consisting of a total of 15 single-choice questions.

\textbf{Part I - Demographics.}

Q1: How old are you?
[18-24], [25-34], [35-44], [45-54], [55-64], [65+], [Prefer not to say]

Q2: What gender do you best identify with?
[Male], [Female], [Non-binary], [Prefer not to say]

Q3: What is the highest level of education you have completed or are currently pursuing?
[High school], [Bachelor's], [Master's], [PhD], [Other], [Prefer not to say]

Q4: What is your field of expertise?
[Computer Science], [Engineering], [Business/Finance], [Arts/Humanities], [Social Sciences], [Health Sciences], [Natural Sciences], [Other], [Prefer not to say]

Q5: How familiar are you with the technique of image-to-video generation?
[Not familiar at all - I have never heard of it], [Not very familiar - I have heard of it but don't know much about it], [Somewhat familiar - I know the basics], [Familiar - I have a good understanding], [Very familiar - I am highly knowledgeable], [Prefer not to say]

\textbf{Part II - Video Selection.}

Q6 - Q15: Please select the video (A, B, C, D, or E) that you believe shows the most significant differences in facial features compared to the reference video (labeled `REF' in the top left corner). Your choice should prioritize the video that best ensures the protection of portrait rights and privacy. (\textit{Note: These videos are in GIF format and can be played by clicking.}) 
[A], [B], [C], [D], [E]

\begin{figure}[htbp]
   \centering
   \includegraphics[width=0.409\textwidth]{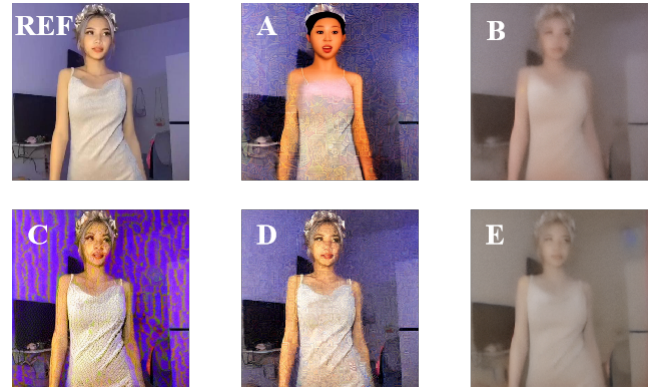}
    \caption{Example of the video selection question.}
    \label{fig:example of the video selection question}
\end{figure}

\subsection{Prompt in GPT-4o Study}
\label{subsec:prompt in gpt-4o study}

We detail our query prompt used in the GPT-4o study. The study is conducted on ChatGPT Web via a browser, using a ChatGPT Plus account. Frame sequences are converted into image grids in PNG format and uploaded via ``Attach files''.

\begin{tcolorbox}[title=Query Prompt to GPT-4o, boxsep=0.92mm]
You will be given six frame sequences: A, B, C, D, E, and a reference frame sequence (REF). Your goal is to rank the quality of the five frame sequences (A, B, C, D, E) from most to least different compared to the reference frame sequence (REF). Base your ranking on how effectively each frame sequence protects the portrait and privacy rights of the individual.

To evaluate the effectiveness of portrait and privacy protection, please consider factors that differentiate each frame sequence from the reference, such as mismatches in the appearance of the human identity, changes in the background, and any additional observable signs. Your evaluation should consider, but not be limited to, these list factors provided merely as examples. Try to explore and assess as many other potential factors as possible.

Please analyze step by step and output your ranking for the five sequences, along with the reasoning which should detail how you evaluate each sequence. You should avoid any potential bias in your evaluation, and ensure that the order in which the frame sequences are presented does not affect your judgment. \\

\noindent Output Format:

Rank: [ordered list from most to least effective]

Reason: [step-by-step analysis]

\begin{center}
   \includegraphics[width=\textwidth]{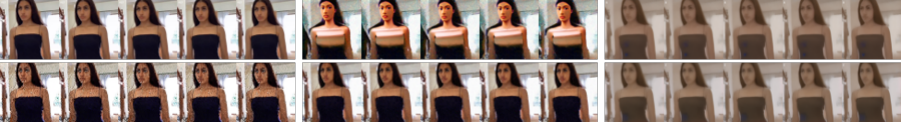}
\end{center}

\end{tcolorbox}

\subsection{Detailed Results}
\label{subsec:detailed results}

\begin{figure}[htbp]
  \centering
  \begin{subfigure}[b]{0.495\linewidth}
        \centering
        \includegraphics[width=\textwidth]{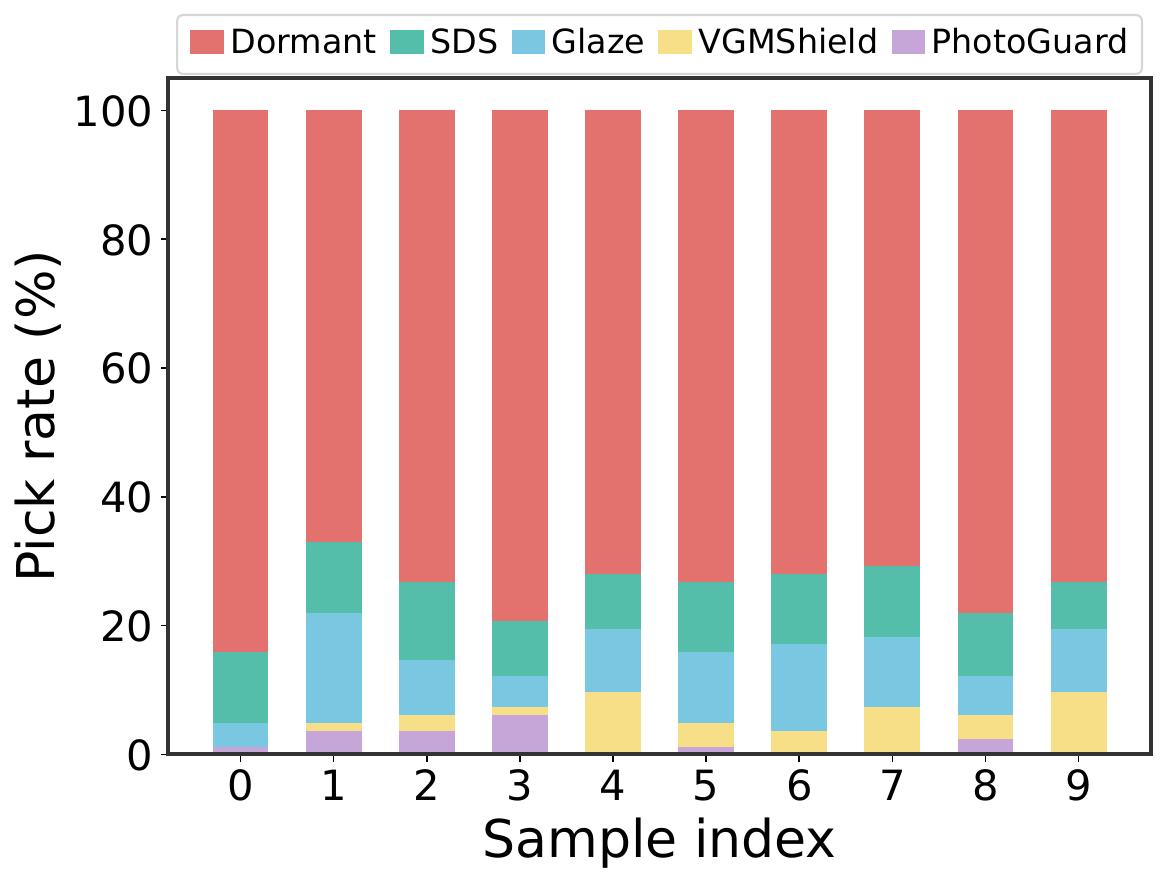}
        \caption{Results of human study}
        \label{fig:results of human study}
  \end{subfigure}
  \hfill
  \begin{subfigure}[b]{0.495\linewidth}
        \centering
        \includegraphics[width=\textwidth]{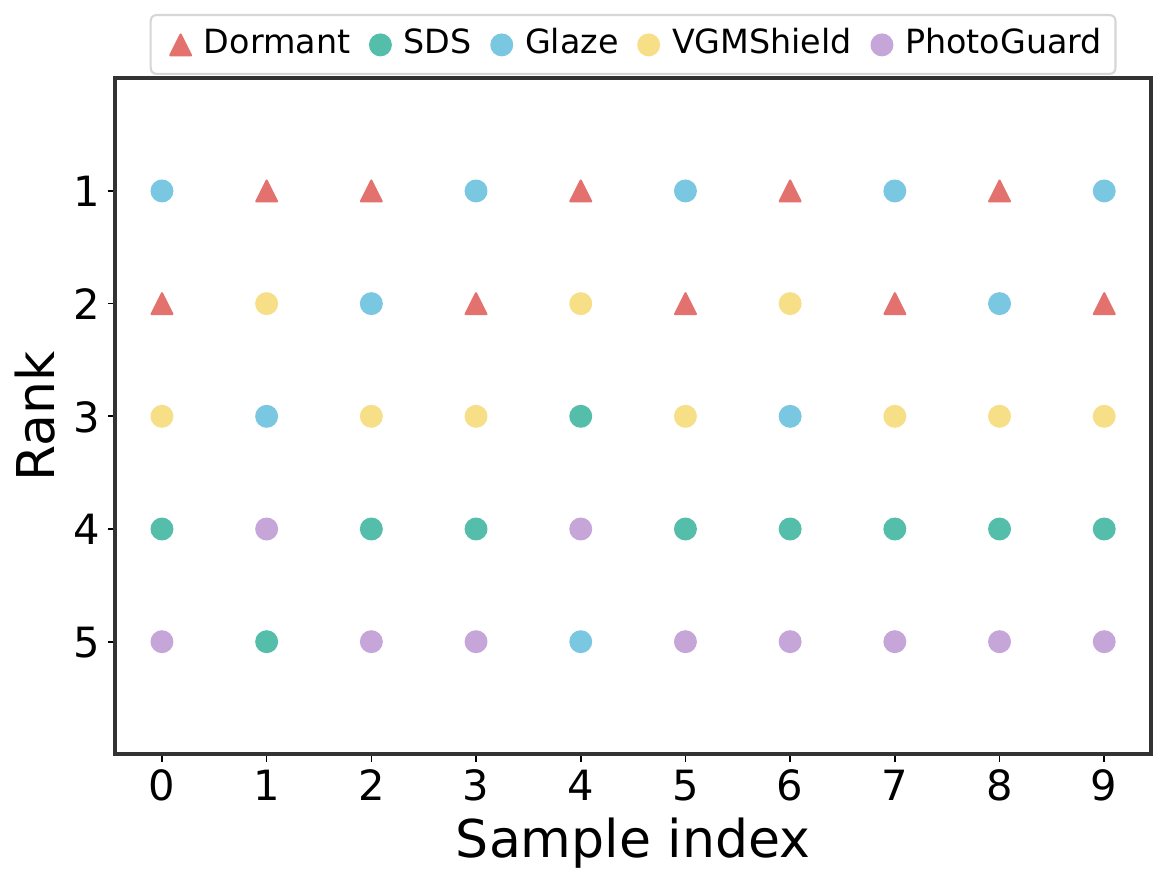}
        \caption{Results of GPT-4o study}
        \label{fig:results of GPT-4o study}
  \end{subfigure}
  \caption{Detailed results of human and GPT-4o studies.}
  \label{fig:results of human and GPT-4o studies}
\end{figure}

Figure~\ref{fig:results of human and GPT-4o studies} presents detailed results for each question in the human and GPT-4o studies. In the human study, \tool~achieves a pick rate of at least 67\% across all samples, with a maximum of 84.15\%. Similarly, \tool~is consistently ranked first or second across all samples by GPT-4o. These results highlight the superiority of \tool~over baseline protections from the perspectives of both human perception and the advanced multimodal large language model.

\section{Image Similarity}
\label{sec:image similarity}

\begin{figure}[htbp]
   \centering
    \includegraphics[width=0.409\textwidth]{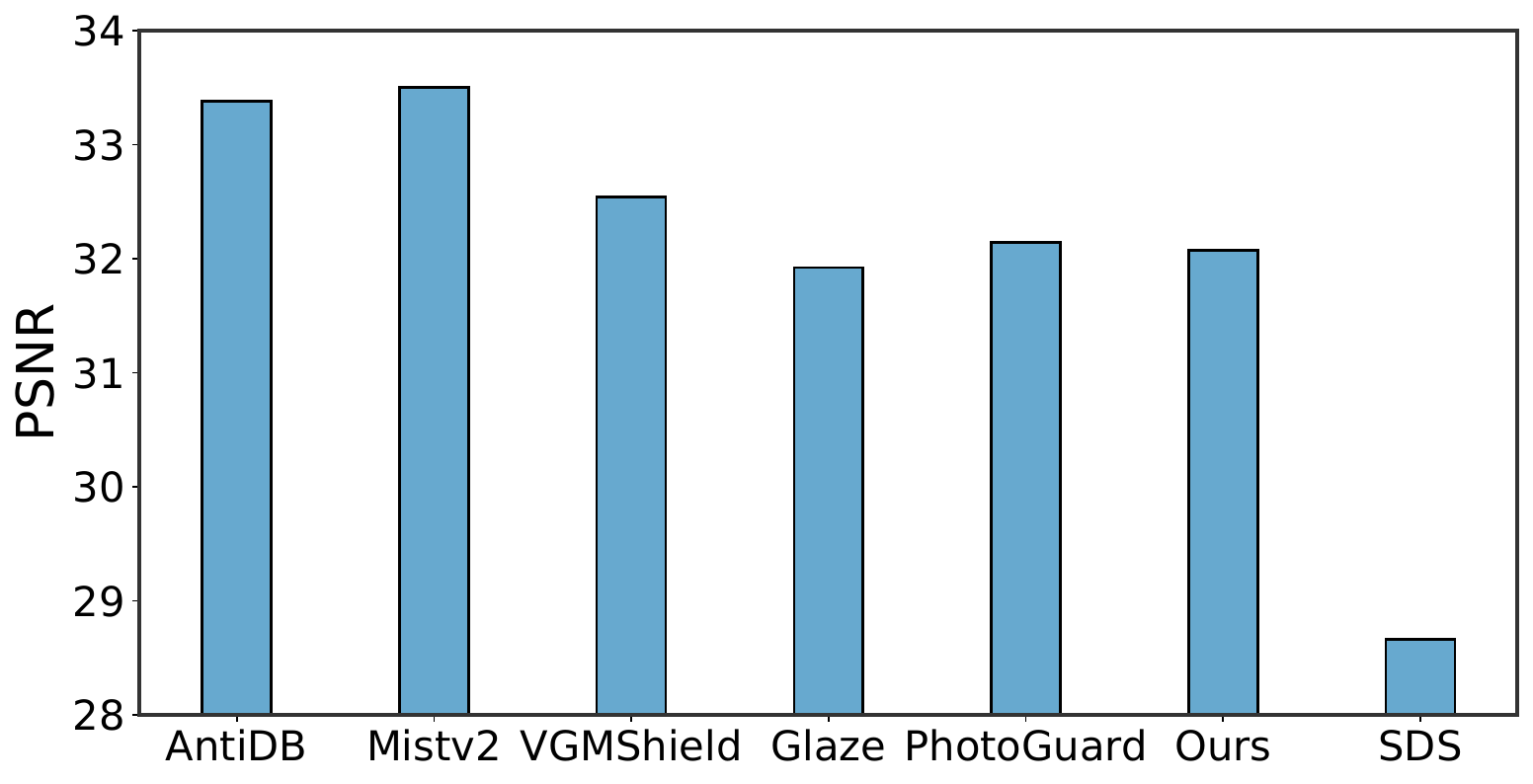}
    \caption{Image similarity before and after protections.}
    \label{fig:image similarity before and after protections}
\end{figure}

We evaluate the similarity between the protected images and their corresponding original images, and present the PSNR for different protection methods in Figure~\ref{fig:image similarity before and after protections}. AntiDB, Mistv2, and VGMShield offer better invisibility but with limited effectiveness in Table~\ref{tab:quantitative comparisons with baseline protections}; Glaze, PhotoGuard, and \tool~exhibit comparable invisibility; and SDS shows the lowest invisibility. While these similarities are computed under the perturbation budget $\eta$ of $16/255$, users can customize $\eta$ to strike a balance between invisibility and protection effectiveness based on their needs, as mentioned in Section~\ref{subsec:ablation study}.

\section{More Visualized Results of Videos}
\label{sec:more visualized results of videos}

\begin{figure*}[!t]
\centering
\includegraphics[width=1\textwidth]{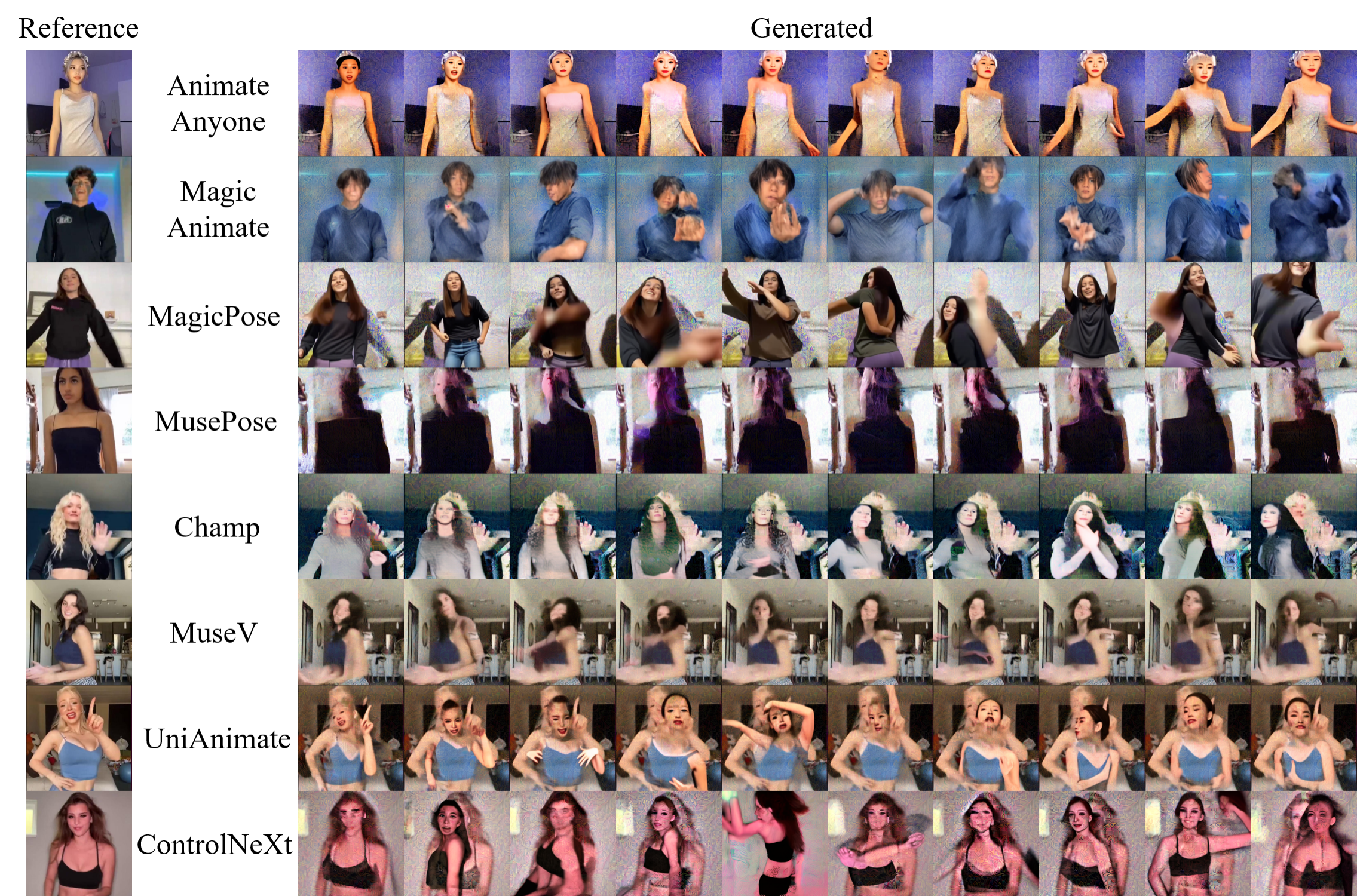}
\caption{More visualization results for videos of \tool~against various pose-driven human image animation methods.}
\label{fig:more visualizations of method}
\end{figure*}

\begin{figure*}[!t]
\centering
\includegraphics[width=1\textwidth]{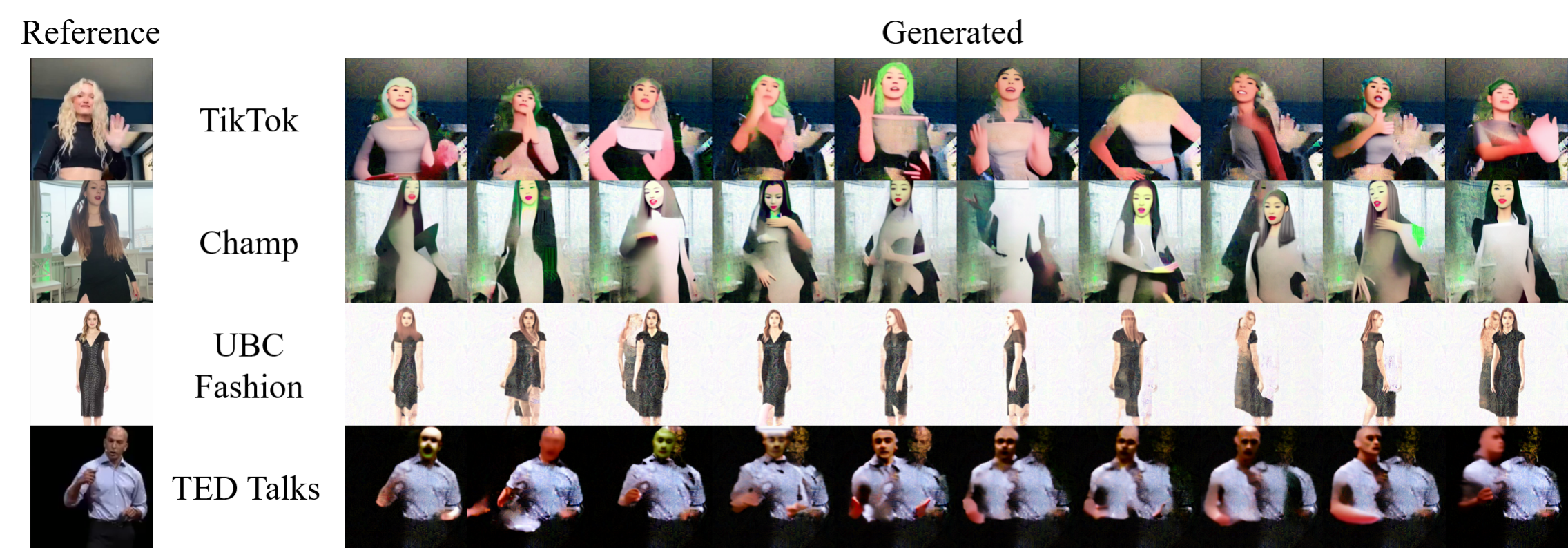}
\caption{More visualization results for videos of \tool~on various datasets.}
\label{fig:more visualizations of dataset}
\end{figure*}

We present ten generated video frames from various pose-driven human image animation methods in Figure~\ref{fig:more visualizations of method} and various datasets in Figure~\ref{fig:more visualizations of dataset}. These visualized results further highlight the effectiveness of \tool, showcasing awful quality, such as misaligned identities, visual distortions, noticeable artifacts, and inconsistent frames.

\end{document}